\documentclass[article,nojss]{jss}

\usepackage[flushleft]{threeparttable}
\usepackage{multirow}
\usepackage{array}
\usepackage{amsmath}
\usepackage{makecell}
\usepackage{float}

\usepackage{float}
\usepackage{placeins}

\usepackage[toc, page]{appendix}
\usepackage{afterpage}
\usepackage{esvect}
\usepackage{ragged2e}
\justifying\let\raggedright\justifying
\usepackage{enumitem}
\usepackage[nodisplayskipstretch]{setspace}
\usepackage{subcaption}
\usepackage{booktabs}
\usepackage{graphicx}
\usepackage{standalone}
\usepackage{amssymb}

\newcommand{\figurehere}[1]{\begin{center}%
		=========================\\%
		Insert Figure #1 about here\\%
		=========================\\%
\end{center}}
\newcommand{\tablehere}[1]{\begin{center}%
		=========================\\%
		Insert Table #1 about here\\%
		=========================\\%
\end{center}}

\usepackage{array}
\newcommand{\PreserveBackslash}[1]{\let\temp=\\#1\let\\=\temp}
\newcolumntype{C}[1]{>{\PreserveBackslash\centering}p{#1}}
\newcolumntype{R}[1]{>{\PreserveBackslash\raggedleft}p{#1}}
\newcolumntype{L}[1]{>{\PreserveBackslash\raggedright}p{#1}}

\author{Jin Liu\\Data Sciences Institute, Takeda Pharmaceuticals}
\Plainauthor{Jin Liu}

\title{Examination of Nonlinear Longitudinal Processes with Latent Variables, Latent Processes, Latent Changes, and Latent Classes in the Structural Equation Modeling Framework: The R package \pkg{nlpsem}}
\Plaintitle{EExamination of Nonlinear Longitudinal Processes with Latent Variables, Latent Processes, Latent Changes, and Latent Classes in the Structural Equation Modeling Framework: The R package \pkg{nlpsem}}
\Shorttitle{\pkg{nlpsem}: Examination of Nonlinear Longitudinal Processes}

\Abstract{
We introduce the R package \pkg{nlpsem}, a comprehensive toolkit for analyzing longitudinal processes within the structural equation modeling (SEM) framework, incorporating individual measurement occasions. This package emphasizes nonlinear longitudinal models, especially intrinsic ones, across four key scenarios: (1) univariate longitudinal processes with latent variables, optionally including covariates such as time-invariant covariates (TICs) and time-varying covariates (TVCs); (2) multivariate longitudinal analyses to explore correlations or unidirectional relationships between longitudinal variables; (3) multiple-group frameworks for comparing manifest classes in scenarios (1) and (2); and (4) mixture models for scenarios (1) and (2), accommodating latent class heterogeneity. Built on the \pkg{OpenMx} R package, \pkg{nlpsem} supports flexible model designs and uses the full information maximum likelihood method for parameter estimation. A notable feature is its algorithm for determining initial values directly from raw data, enhancing computational efficiency and convergence. Furthermore, \pkg{nlpsem} provides tools for goodness-of-fit tests, cluster analyses, visualization, derivation of p-values and three types of confidence intervals, as well as model selection for nested models using likelihood ratio tests and for non-nested models based on criteria such as Akaike Information Criterion and Bayesian Information Criterion. This article serves as a companion document to the \pkg{nlpsem} R package, providing a comprehensive guide to its modeling capabilities, estimation methods, implementation features, and application examples using synthetic intelligence growth data.}

\Plainkeywords{Latent Growth Curve Models and Latent Change Score Models; Multivariate Growth Models and Longitudinal Mediation Model; Multiple-group Growth Modeling; Growth Mixture Modeling; Time-invariant Covariates and Time-varying Covariates; Individual Measurement Occasions}

\Address{
  Jin Liu \\
  E-mail: \email{Veronica.Liu0206@gmail.com}
}

\usepackage{Sweave}
\begin{document}
\section{Introduction}\label{sec:intro}
Longitudinal data are prevalent in fields like psychology, behavioral sciences, biomedicine, and social sciences, where researchers aim to understand how individuals change over time. Capturing the complex patterns of change inherent in such data necessitates specialized statistical methods capable of modeling individual trajectories and evolving relationships. Advanced statistical models enable researchers to decode these complexities, providing insights into developmental processes, treatment effects, and other temporal phenomena.

Two primary frameworks dominate the analysis of longitudinal data: mixed-effects modeling and structural equation modeling (SEM). In SEM, latent variables known as growth factors are used to capture underlying change patterns, with the means and variances representing the average trajectory and individual deviations, respectively. Mixed-effects models provide a parallel approach, where fixed effects correspond to the means of growth factors and random effects correspond to their variances. Both frameworks have been shown to be theoretically and empirically equivalent in certain contexts \citep{Bauer2003Estimating, Curran2003Multilevel}.

While linear longitudinal models—typically involving two latent variables, intercept and slope—are widely used and often suffice for short-term studies, they may fall short in capturing the nuances of data collected over extended periods or with frequent measurements. Nonlinear longitudinal models are particularly suited to analyzing complex change patterns. The need for robust tools that can handle such nonlinearities within a flexible framework has become increasingly apparent.

\subsection{Exploring Nonlinear Longitudinal Analysis: Challenges and Existing Computational Tools}\label{intro:basic}
Nonlinear longitudinal models can be classified based on their nonlinearity concerning (1) time, (2) parameters, and (3) growth factors \citep[Chapter~9]{Grimm2016growth}. The third category, known as intrinsically nonlinear, is particularly intricate, as these models involve derivatives with respect to one growth factor that depend on another. These models necessitate multidimensional integrations of joint likelihood over growth factors and lack straightforward closed-form likelihood function \citep{Rohloff2022nonlinear}. Addressing these complexities requires approximation techniques within both frameworks.

Various computational tools exist for analyzing longitudinal data, particularly for models exhibiting nonlinearity in time or parameters. Within the mixed-effects modeling framework, packages like \pkg{lme4} \citep{Bates2014lme4, Bates2015lme4}, \pkg{nlme} \citep{Pinheiro2000nlme, Pinheiro2023nlme}, and \pkg{lcmm} \citep{Proust2017lcmm} are noteworthy. While \pkg{lme4} and \pkg{nlme} are recognized for their robust functionality, \pkg{lcmm} excels in fitting joint and mixture models. In the SEM framework, tools such as \pkg{OpenMx} \citep{OpenMx2016package, Pritikin2015OpenMx, Hunter2018OpenMx, User2020OpenMx}, \pkg{lavaan} \citep{rosseel2012lavaan}, and \pkg{Mplus} \citep{Muthen2017Mplus}, provide broader flexibility but require careful model specification and validation. However, these tools have limitations in handling intrinsically nonlinear models, especially in the framework of individual measurement occasions.

Analyzing intrinsically nonlinear models often requires specialized approximation techniques. In mixed-effects modeling, marginal maximum likelihood estimation (MLE) methods are commonly used \citep{Harring2006nonlinear, Toit2009NRC, Cudeck2010nonlinear}, while in SEM, Taylor series expansion is often employed \citep{Browne1991Taylor, Preacher2015repara}. Some R routines, such as \pkg{fitPMM} \citep{Zopluoglu2014knot}, implement marginal MLE for specific functional forms but are limited in flexibility, particularly regarding the inclusion of covariates.

Additionally, addressing unstructured measurement schedules adds another layer of complexity. In many research settings, such as clinical trials or longitudinal surveys, individuals may be assessed at varying time points, leading to irregular measurement occasions. While the mixed-effects modeling framework addresses this by treating measurement time as a continuous variable, the SEM framework faces challenges due to its reliance on wide-format data structures. A common approach is to divide the assessment period into multiple time intervals (time-windows), allowing one response per interval per participant. However, this approach has limitations, as ignoring individual time variations can result in biased estimates \citep{Blozis2008coding, Coulombe2015ignoring}. Techniques like the `definition variables' approach \citep{Mehta2000people, Mehta2005people} support the inclusion of individual-specific measurement times but can be challenging to implement in SEM software.

To address these limitations, we introduce the \pkg{nlpsem} package, which provides a comprehensive set of tools specifically designed for intrinsically nonlinear LGCMs within the SEM framework. The current version of \pkg{nlpsem} (v0.3) focuses on commonly used and practically important functional forms, such as the negative exponential function for individual growth rate ratios \citep{Sterba2014individually}, the Jenss-Bayley function for individual growth acceleration ratios \citep[Chapter~12]{Grimm2016growth}, and the bilinear spline function (or linear-linear piecewise function) for estimating individual knots \citep{Liu2019BLSGM}. Additionally, \pkg{nlpsem} provides a parsimonious version of each intrinsically nonlinear LGCM, supports the estimation of models with quadratic functional forms, and accommodates linear longitudinal models, making it versatile for a wide range of research applications. By seamlessly incorporating individual measurement occasions through the `definition variables' technique, \pkg{nlpsem} simplifies the analytical process for researchers working with unstructured measurement schedules. Table \ref{tbl:compare1} provides a comparison of the advantages and limitations of \pkg{nlpsem} relative to other tools in this field.

\tablehere{1}

\subsection{Addressing Advanced Longitudinal Challenges: The nlpsem Approach}\label{intro:adv}
Beyond modeling nonlinear longitudinal trajectories, researchers often face advanced research questions and complex data patterns in longitudinal studies. These challenges extend from evaluating time-dependent states to assessing time-dependent changes, such as interval-specific slopes, interval-specific changes, and changes from baseline. The complexity further increases when multiple longitudinal variables are analyzed simultaneously, requiring an understanding of their interconnected dynamics, whether through basic correlations or unidirectional relationships. While packages like \pkg{lcmm} can capture correlations between multiple outcomes, they are limited in analyzing unidirectional relationships—a gap that \pkg{nlpsem} addresses effectively. By leveraging the versatility of SEM, \pkg{nlpsem} accommodates nonlinear trajectories and the intricate dynamics of multiple longitudinal variables, offering a holistic perspective that extends beyond the capabilities of mixed-effects models. Within this framework, \pkg{nlpsem} provides:
\begin{enumerate}
\item \textbf{Rate of Change Estimation Over Time:} Probing into both intrinsic and nonintrinsic nonlinear processes \citep{Liu2022LCSM, Grimm2013LCSM1}.
\item \textbf{Influence of Baseline Characteristics on Trajectories:} Evaluating the impact on longitudinal trajectories of baseline characteristics, as detailed in \citet[Chapter~5]{Grimm2016growth} and \citet{Liu2019BLSGM}.
\item \textbf{Comprehensive Analysis of Joint Processes,} which includes:
\begin{itemize}
\item \textbf{Effect of Time-Varying Covariates (TVC) on Outcomes:} Analyzing instances where one process serves as a longitudinal outcome and another as a TVC, as discussed in \citet{Grimm2007multi}, \citet[Chapter~8]{Grimm2016growth}, \citet{Liu2022decompose} and \citet{Liu2023GMMTVC}.
\item \textbf{Correlation Among Multiple Longitudinal Processes:} Examining situations where all processes are longitudinal outcomes, exploring their interconnections, highlighted in \citet{Blozis2004MGM}, \citet{Ferrer2003MGM}, \citet[Chapter~8]{Grimm2016growth}, and \citet{Liu2021PBLSGM}.
\item \textbf{Direct and Mediated Effects of Predictors on Outcomes:} Assessing how a predictor influences a longitudinal outcome, both directly and via a mediator \citep{Cheong2003mediate, Soest2011mediate, MacKinnon2008mediate, Liu2022mediate}.
\end{itemize}
\item \textbf{Change Patterns Across Diverse Groups:} Investigating the longitudinal processes unique to different groups over time.
\item \textbf{Discovery of Latent Classes:} Identifying latent classes and understanding baseline characteristics that inform these classes and explains the variability within each group when applicable, as explored in \citet{Liu2021MoE}.
\end{enumerate}

Overall, \pkg{nlpsem} supports complex longitudinal analyses, accommodating multivariate growth models, time-varying covariates, mediation, and multiple group and mixture models through the `definition variables' approach.

\subsection{Scope and Focus of nlpsem}\label{I:scope}
To clarify the scope and focus of the \pkg{nlpsem} package, there are multiple approaches to analyzing longitudinal data, each catering to different research needs and objectives. The \pkg{nlpsem} package is specifically designed to model trajectories by defining change patterns with functional forms, rather than exploring hidden Markov latent variables \citep{Song2017Hidden} or dynamic latent factor models \citep{Zhang2024dynamic} that are often used in time series or other longitudinal data analyses. These areas are distinct in their treatment of non-linearity and are not covered by \pkg{nlpsem}. 

In its current version, \pkg{nlpsem} supports both linear and several nonlinear functional forms—such as the negative exponential, Jenss-Bayley, and bilinear spline functions—which are among the most commonly used in the field. While the package does not encompass all nonlinear functional forms, such as logistic and Gompertz functions, it is tailored to meet the needs of a broad research community. Additional functional forms could be incorporated in future updates based on user demand and evolving research practices.

The remainder of this paper is organized as follows. Section \ref{sec:method} presents the theoretical foundation for nonlinear longitudinal models within the SEM framework, including model specification, estimation techniques, initial value setup, model optimization, methods for deriving non-estimable parameters, and post-fit computations. Section \ref{sec:imp} provides a detailed overview of the \pkg{nlpsem} implementation, with guidance on using estimation functions and specifying models. Section \ref{sec:example} demonstrates \pkg{nlpsem}’s functionality through applied examples on real-world datasets, with complete code and output accessible online. Finally, Section \ref{sec:conclude} discusses practical insights, limitations, and potential future directions for nonlinear longitudinal modeling with \pkg{nlpsem}.

\section{Methodological Framework of the nlpsem Package}\label{sec:method}
This section outlines the core methodological components for modeling nonlinear longitudinal data within the SEM framework. It focuses on the model specification of key model families and their associated estimation methods, providing a theoretical foundation for analyzing complex longitudinal processes.

\subsection{Model Specification}\label{M:spec}
In this section, we provide detailed specifications of the modeling families implemented in the \pkg{nlpsem} package. We begin by introducing univariate longitudinal models, specifically latent growth curve models (LGCMs) and latent change score models (LCSMs), both with and without time-invariant covariates (TICs). We then explore multivariate longitudinal processes, including LGCMs or LCSMs with time-varying covariates (TVCs), multivariate growth models, and longitudinal mediation models. Finally, we address multiple-group and mixture models for analyzing heterogeneous populations with observed and latent classes.

\subsubsection{Latent Growth Curve Models}\label{spec:LGCMs}
The Latent Growth Curve Model (LGCM), within the SEM framework, is widely used to analyze growth over time. This subsection provides an overview of LGCMs, focusing on three intrinsically nonlinear functional forms, their simplified versions, and two other commonly used forms: linear and quadratic curves. The general form of an LGCM with individual measurement occasions is
\begin{equation}\label{eq:LGCM1}
\boldsymbol{y}_{i}=\boldsymbol{\Lambda}_{i}\times\boldsymbol{\eta}_{i}+\boldsymbol{\epsilon}_{i},
\end{equation}
where $\boldsymbol{y}_{i}$ is a $J\times1$ vector of repeated measurements for individual $i$, with $J$ being the number of measurements. The vector $\boldsymbol{\eta}_{i}$ ($K\times1$) contains growth factors, which are latent variables representing the growth status of individual $i$, with $K$ being the number of growth factors. Furthermore, $\boldsymbol{\Lambda}_{i}$ is a $J\times K$ matrix of factor loadings, dependent on individual measurement times. The vector $\boldsymbol{\epsilon}_{i}$ ($J\times1$) represents residuals, assumed to follow a multivariate normal distribution (i.e., $\boldsymbol{\epsilon}_{i}\sim\text{MVN}\big(\boldsymbol{0},\theta_{\epsilon}\boldsymbol{I}\big)$), where $\theta_{\epsilon}$ is the residual variance and $\boldsymbol{I}$ is a $J\times J$ identity matrix. The growth factors can be expressed as deviations from their mean values
\begin{equation}\label{eq:LGCM2_noTIC}
\boldsymbol{\eta}_{i}=\boldsymbol{\mu_{\eta}}+\boldsymbol{\zeta}_{i},
\end{equation}
where $\boldsymbol{\mu_{\eta}}$ is a $K\times1$ vector of growth factor means, and $\boldsymbol{\zeta}_{i}$ is a $K\times1$ vector representing individual deviations from these means. To assess the impact of time-invariant covariates (TICs) on growth factors and curves, the growth factors can be regressed on the covariates,
\begin{equation}\label{eq:LGCM2_TIC}
\boldsymbol{\eta}_{i}=\boldsymbol{\alpha}+\boldsymbol{B}_{\text{TIC}}\times\boldsymbol{X}_{i}+\boldsymbol{\zeta}_{i},
\end{equation}
where $\boldsymbol{\alpha}$ is a $K\times1$ vector of growth factor intercepts, $\boldsymbol{B}_{\text{TIC}}$ is a $K\times m$ matrix of regression coefficients from TICs to growth factors (with $m$ representing the number of TICs), and $\boldsymbol{X}_{i}$ is an $m\times1$ vector of TICs for individual $i$. Growth factors in Equation \ref{eq:LGCM1} are assumed to follow a (conditional) multivariate normal distribution: $\boldsymbol{\zeta}_{i}\sim \text{MVN}\big(\boldsymbol{0}, \boldsymbol{\Psi}_{\boldsymbol{\eta}}\big)$, where $\boldsymbol{\Psi}_{\boldsymbol{\eta}}$ is a $K\times K$ matrix representing the variance and unexplained variance of growth factors representing the variance and unexplained variance of growth factors without and with TICs, respectively.

Table \ref{tbl:LGCM_summary} provides the model specification of LGCM with multiple commonly used functional forms and corresponding interpretation of growth coefficients. For example, in the linear functional form, the simplest model, there are two key coefficients: the intercept ($\eta_{0i}$) and the linear slope ($\eta_{1i}$). These parameters vary across individuals, and understanding these variations is central to analyzing between-individual differences in within-individual changes over time \citep{Biesanz2004Linear, Zhang2012LCSM, Grimm2013LCSM1}.

\tablehere{2}

While linear function coefficients are straightforward to interpret, they are insufficient for capturing extended, nonlinear change patterns. To address this limitation, more complex functional forms with additional growth coefficients are necessary, particularly for analyzing long-term trajectories. For instance, a quadratic functional form, which includes an acceleration growth factor ($\eta_{2i}$) in addition to the intercept ($\eta_{0i}$) and linear slope ($\eta_{1i}$), is suitable for describing nonlinear trajectories, particularly when growth acceleration is of interest. LGCMs with a quadratic form are Type I nonlinear models, where factor loadings depend only on measurement occasions. Other nonlinear functions in Table \ref{tbl:LGCM_summary} also provide valuable insights. For example, $\eta_{1i}$ in a negative exponential function reflects an individual's growth capacity.

As shown in Table \ref{tbl:LGCM_summary}, the second term of the individual growth curve for the negative exponential function involves two growth factors, $\eta_{1i}$ and $b_{i}$, making the model intrinsically nonlinear since the derivatives of the growth curve with respect to $\eta_{1i}$ or $b_{i}$ are functions of each other. Consequently, the negative exponential function with individual coefficient $b_{i}$ cannot be directly modeled within the SEM framework because the factor loadings matrix cannot depend on a growth factor. To address this, two approaches are commonly used: (1) linearizing the function using Taylor series expansion \citep{Preacher2015repara}, or (2) employing a reduced model that assumes a consistent growth rate ratio across the population. These methods can also be applied to handle $c_{i}$ in the Jenss-Bayley function and $\gamma_{i}$ in the bilinear spline function with an unknown knot. After linearization, the factor-loading matrices for these intrinsically nonlinear models depend on the growth factor means and measurement occasions, making them estimable within the SEM framework. Technical details on linearization for these functions are provided by \citet{Sterba2014individually}, \citet[Chapter~12]{Grimm2016growth}, and \citet{Liu2019BLSGM}. The reduced versions of these models are classified as Type II nonlinear models, where factor loading matrices depend on population-level coefficients $b$, $c$, or $\gamma$ in addition to measurement times. 

Of the nonlinear functional forms discussed, the bilinear spline (i.e., linear-linear piecewise) with an unknown knot presents additional complexities for SEM implementation. Unlike other forms, it requires a uniform expression for SEM implementation. Multiple reparameterization techniques are available to create a uniform expression before and after the knot (i.e., $\gamma_{i}$, or $\gamma$ in the reduced model). For further details, see \citet{Harring2006nonlinear}, \citet[Chapter~11]{Grimm2016growth}, and \citet{Liu2019BLSGM}. 

\subsubsection{Latent Change Score Models}\label{spec:LCSMs}
The primary purpose of using LGCMs is to characterize time-dependent states. However, when exploring nonlinear longitudinal processes, one may also be interested in assessing the growth rate (i.e., rate-of-change) \citep{Grimm2013LCSM1, Grimm2013LCSM2, Zhang2012LCSM} and the cumulative value of the growth rate over time \citep{Liu2022LCSM}. In such scenarios, LCSMs, which emphasize time-dependent changes, are more appropriate. LCSMs, also known as latent difference score models \citep{McArdle2001LCSM1, McArdle2001LCSM2, McArdle2009LCSM}, were developed to incorporate difference equations with discrete measurement occasions into the SEM framework. Recent advancements by \citet{Liu2022JB_LCSM, Liu2021PLBGM, Liu2022LCSM} introduced a novel specification for LCSMs that accommodates unequal study waves and individual measurement occasions using the `definition variables' approach. These works provide the theoretical foundation for LCSMs. The model specification of LCSMs begins with the concept of classical test theory
\begin{equation}\label{eq:LCSM1}
y_{ij}=y^{*}_{ij}+\epsilon_{ij},
\end{equation}
where $y_{ij}$, $y^{*}_{ij}$, and $\epsilon_{ij}$ represent the observed score, latent true score, and residual for individual $i$ at time $j$, respectively. This specification indicates that an individual's observed score at a given time can be decomposed into a latent true score and a residual. At baseline (i.e., $j=1$), the true score reflects the growth factor representing the initial status. For each subsequent time point (i.e., $j\ge2$), the true score at time $j$ is a linear combination of the previous score at $j-1$ and the true change that occurs between time $j-1$ and $j$
\begin{equation}\label{eq:LCSM2}
y^{*}_{ij}=\begin{cases}
\eta_{0i}, & \text{if $j=1$}\\
y^{*}_{i(j-1)}+\delta y_{ij}, & \text{if $j=2, \dots, J$}
\end{cases},
\end{equation}
where $\delta y_{ij}$ represents the change during the $(j-1)^{th}$ interval (i.e., from $j-1$ to $j$) for the $i^{th}$ individual. These interval-specific changes can be expressed as the product of the interval-specific slopes and the time interval. \citet{Liu2022LCSM} identified two methods to express these slopes based on research interests. When the focus is on assessing growth rates over time, a process with $J$ measurements can be viewed as a linear piecewise function with $J-1$ segments (i.e., a nonparametric functional form)
\begin{align}
&\delta y_{ij}=dy_{ij}\times(t_{ij}-t_{i(j-1)})\qquad (j=2, \dots, J), \label{eq:LCSM3_nonpara}\\
&dy_{ij}=\eta_{1i}\times\gamma_{j-1}\qquad (j=2, \dots, J). \label{eq:LCSM4_nonpara}
\end{align}
In Equation \ref{eq:LCSM3_nonpara}, the interval-specific changes are defined as the product of interval-specific slopes ($dy_{ij}$) and the corresponding time intervals, and $dy_{ij}$ is further defined in Equation \ref{eq:LCSM4_nonpara} as the product of the shape factor for the first interval slope and the relative rate.

If additional features, such as growth acceleration or capacity, are of interest, a parametric nonlinear function (e.g., quadratic, negative exponential, Jenss-Bayley) may be used. In this case, the slope within an interval is not constant. \citet{Liu2022LCSM} and \citet{Liu2022JB_LCSM} suggest approximating interval-specific changes as the product of the instantaneous midpoint slope and the interval length
\begin{equation}\label{eq:LCSM3_para}
\delta y_{ij}\approx dy_{ij\_mid}\times(t_{ij}-t_{i(j-1)}), 
\end{equation}
where $dy_{ij\_mid}$ is the instantaneous slope at the midpoint of the $(j-1)^{th}$ interval. Table \ref{tbl:LCSM_summary} provides the expression of $dy_{ij}$ for the linear piecewise function and $dy_{ij_mid}$ for quadratic, negative exponential, and Jenss-Bayley functions. The LCSMs can also accommodate intrinsically nonlinear forms. Specifically, \citet{Grimm2013LCSM1} and \citet{Liu2022JB_LCSM} introduced LCSMs for the negative exponential function with a random growth rate ratio and the Jenss-Bayley function with a random growth acceleration ratio. Established methods enable the implementation of intrinsically nonlinear LCSMs, their reduced models, and LCSMs with linear piecewise or quadratic forms.

\tablehere{3}

LCSMs can be expressed in matrix form, similar to Equation \ref{eq:LGCM1}. They can further be represented as Equations \ref{eq:LGCM2_noTIC} and \ref{eq:LGCM2_TIC}, depending on whether TICs are included. The distribution assumptions for LCSMs are the same as those for LGCMs. Table \ref{tbl:LCSM_summary} outlines the growth factors and corresponding factor loadings for each model. The interpretation of growth factors in LCSMs is consistent with LGCMs, with the first element in vector $\boldsymbol{\eta}_{i}$ representing the initial status and the remaining elements representing the growth rate. Unlike LGCMs, LCSMs focus on accumulated change, so their factor loadings are based on interval-specific growth rates and time intervals, where the time intervals act as `definition variables' in models with unstructured time frames.

A key advantage of LCSMs is their ability to evaluate interval-specific slopes, changes, and amounts of change from baseline. Two methods can be used for this evaluation. The first involves deriving expressions for the means and variances of these parameters, treating them as non-estimable. Table \ref{tbl:LCSM_summary}
provides these expressions. Alternatively, when applicable, interval-specific slopes, changes, and amounts of change from baseline can be defined as additional latent variables, allowing for post-fit calculation of factor scores.

\subsubsection{Longitudinal Models with Time-varying Covariates}\label{spec:TVC}
A key interest in evaluating longitudinal processes is understanding how a covariate affects between-individual differences in within-individual change. As discussed earlier, TICs can be included in LGCMs or LCSMs to account for variability in growth factors and curves. However, covariates in longitudinal studies are not always TICs. \citet{Grimm2007multi} proposed using LGCMs with a time-varying covariate (TVC) to analyze bivariate longitudinal variables, treating the primary process as the outcome and the other variable as the TVC. Table \ref{tbl:TVC_summary} presents the model specifications. Recent research, including \citet[Chapter~8]{Grimm2016growth}, \citet{Liu2022decompose}, and \citet{Liu2023GMMTVC}, has explored the advantages and drawbacks of this approach. To address limitations in the model proposed by \citet{Grimm2007multi}, \citet{Liu2022decompose} developed three methods to decompose a TVC into an initial trait and temporal states, enabling separate evaluation of baseline and temporal effects on outcome trajectories. These methods are based on the LCSM specification with a piecewise linear function introduced in Subsection \ref{spec:LCSMs}. Table \ref{tbl:TVC_summary} provides the technical details for these decomposition methods.

\tablehere{4}

Across all three decomposition methods, the initial trait remains consistent and is represented by the true score of the baseline value, indicating the TVC's initial status (i.e., $\eta^{[x]}_{0i}$ in Table \ref{tbl:TVC_summary}). The temporal states, however, vary across methods. Specifically, interval-specific slopes (i.e., $\boldsymbol{dx}_{i}$ in Table \ref{tbl:TVC_summary}), interval-specific changes (i.e., $\boldsymbol{\delta x}_{i}$ in Table \ref{tbl:TVC_summary}), and change-from-baseline amounts (i.e., $\boldsymbol{\Delta x}_{i}$ in Table \ref{tbl:TVC_summary}) are used to characterize each approach. Regressing the longitudinal outcome's growth factors on the initial trait evaluates the baseline effect (i.e., $\boldsymbol{\beta}_{\text{TVC}}$ in Table \ref{tbl:TVC_summary}), while regressing each post-baseline observed measurement on the corresponding temporal state assesses the temporal effect (i.e., $\kappa$ in Table \ref{tbl:TVC_summary}). Each longitudinal model introduced in Sections \ref{spec:LGCMs} and \ref{spec:LCSMs}, can incorporate a TVC to account for dynamic influences over time. More technical details on these methods and their applications can be found in \citet{Liu2022decompose} and \citet{Grimm2007multi}.

\subsubsection{Parallel Processes and Correlated Growth Models}\label{spec:MGMs}
All processes under investigation can be viewed as longitudinal outcomes and analyzed using multivariate growth models (MGMs) \citep[Chapter~8]{Grimm2016growth}, also known as parallel process or correlated growth models \citep{McArdle1988Multi, Grimm2007multi}. MGMs examine covariances among cross-process growth factors, capturing relationships across multiple longitudinal processes. A general specification for a bivariate growth model is
\begin{equation}\label{eq:MGMs1}
\begin{pmatrix}
\boldsymbol{y}_{i} \\ \boldsymbol{z}_{i}
\end{pmatrix}=
\begin{pmatrix}
\boldsymbol{\Lambda}_{i}^{[y]} & \boldsymbol{0} \\ \boldsymbol{0} & \boldsymbol{\Lambda}_{i}^{[z]}
\end{pmatrix}\times
\begin{pmatrix}
\boldsymbol{\eta}^{[y]}_{i} \\ \boldsymbol{\eta}^{[z]}_{i}
\end{pmatrix}+
\begin{pmatrix}
\boldsymbol{\epsilon}^{[y]}_{i} \\ \boldsymbol{\epsilon}^{[z]}_{i}
\end{pmatrix},
\end{equation}
where $\boldsymbol{z}_{i}$ is a vector of repeated measurements for the second longitudinal outcome of individual $i$. While $\boldsymbol{y}_{i}$ and $\boldsymbol{z}_{i}$ often assume the same functional forms, their time structures may differ. The growth factors can be further expressed as deviations from the their mean values
\begin{equation}\label{eq:MGMs2}
\begin{pmatrix}
\boldsymbol{\eta}^{[y]}_{i} \\ \boldsymbol{\eta}^{[z]}_{i}
\end{pmatrix}=
\begin{pmatrix}
\boldsymbol{\mu}^{[y]}_{\boldsymbol{\eta}} \\ \boldsymbol{\mu}^{[z]}_{\boldsymbol{\eta}}
\end{pmatrix}+
\begin{pmatrix}
\boldsymbol{\zeta}^{[y]}_{i} \\
\boldsymbol{\zeta}^{[z]}_{i}
\end{pmatrix},
\end{equation}
where  $\boldsymbol{\mu}^{[y]}_{\boldsymbol{\eta}}$ and $\boldsymbol{\mu}^{[z]}_{\boldsymbol{\eta}}$ are $K\times1$ vectors of growth factor means, while $\boldsymbol{\zeta}^{[y]}_{i}$ and $\boldsymbol{\zeta}^{[z]}_{i}$ are $K\times1$ vectors of individual deviations from those means. In practice, these individual deviations are often assumed to follow a multivariate normal distribution
\begin{equation}\label{eq:MGMs3}
\begin{pmatrix} 
\boldsymbol{\zeta}^{[y]}_{i} \\ \boldsymbol{\zeta}^{[z]}_{i}
\end{pmatrix}\sim \text{MVN}\bigg(\boldsymbol{0}, 
\begin{pmatrix}
\boldsymbol{\Psi}_{\boldsymbol{\eta}}^{[y]} & \boldsymbol{\Psi}_{\boldsymbol{\eta}}^{[yz]} \\
& \boldsymbol{\Psi}_{\boldsymbol{\eta}}^{[z]}
\end{pmatrix}\bigg),
\end{equation}
where $\boldsymbol{\Psi}_{\boldsymbol{\eta}}^{[y]}$ and $\boldsymbol{\Psi}_{\boldsymbol{\eta}}^{[z]}$ are $K\times K$ variance-covariance matrices for outcome-specific growth factors, and $\boldsymbol{\Psi}_{\boldsymbol{\eta}}^{[yz]}$, a $K\times K$ matrix, represents covariances between cross-outcome growth factors. When $\boldsymbol{y}_{i}$ and $\boldsymbol{z}_{i}$ are linear functions, $\boldsymbol{\Psi}_{\boldsymbol{\eta}}^{[yz]}$ estimates intercept-intercept, slope-slope, and intercept-slope covariances across outcomes. Individual residuals $\boldsymbol{\epsilon}^{[y]}_{i}$ and $\boldsymbol{\epsilon}^{[z]}_{i}$ are often assumed to follow independent normal distributions with homogeneous covariances over time
\begin{equation}\label{eq:MGMs4}
\begin{pmatrix} 
\boldsymbol{\epsilon}^{[y]}_{i} \\ \boldsymbol{\epsilon}^{[z]}_{i}
\end{pmatrix}\sim \text{MVN}\bigg(\boldsymbol{0}, 
\begin{pmatrix}
\theta^{[y]}_{\epsilon}\boldsymbol{I} & \theta^{[yz]}_{\epsilon}\boldsymbol{I} \\
& \theta^{[z]}_{\epsilon}\boldsymbol{I}
\end{pmatrix}\bigg),
\end{equation}
where $\theta^{[y]}_{\epsilon}$ and $\theta^{[z]}_{\epsilon}$ are residual variances of the longitudinal outcomes $\boldsymbol{y}_{i}$ and $\boldsymbol{z}_{i}$, respectively, $\theta^{[yz]}_{\epsilon}$ is the residual covariance between the two processes, and $\boldsymbol{I}$ is a $J\times J$ identity matrix if both $\boldsymbol{y}_{i}$ and $\boldsymbol{z}_{i}$ have $J$ repeated measurements. 

Previous studies have developed MGMs using LGCMs or LCSMs for univariate processes, such as the bilinear spline growth curve model with an unknown knot \citep{Liu2021PBLSGM} and MGMs with quadratic and negative exponential functions \citep{Blozis2004MGM}. Building on these advancements, MGMs can now integrate each univariate longitudinal model discussed in Sections \ref{spec:LGCMs} and \ref{spec:LCSMs} to explore correlation among multiple longitudinal processes, even within the framework of individual measurement occasions.

\subsubsection{Longitudinal Mediation Models}\label{spec:MED}
The third statistical model for multivariate longitudinal processes is the longitudinal mediation model, designed to evaluate how a predictor affects an outcome through direct and indirect (mediated) pathways \citep{Baron1986indirect}. Because mediation relationships often evolve over time, longitudinal data are preferred for testing mediation hypotheses. Previous studies, such as \citet{Hayes2009Mediate, MacKinnon2000Mediator, MacKinnon2002Mediator, Cheung2008Mediate, Shrout2002Mediate, Selig2009mediate, Gollob1987mediate, Cole2003mediate, Maxwell2007mediate, MacKinnon2008mediate, Cheong2003mediate, Soest2011mediate}, highlight the limitations of cross-sectional data and the advantages of longitudinal data in mediation analysis.

\citet{Cheong2003mediate} proposed using the LGCM framework for longitudinal mediation, developing a parallel linear growth model to explore how a baseline predictor influences outcome changes via changes in the mediator. \citet{Soest2011mediate} expanded this model with additional regression paths, and \citet[Chapter~8]{MacKinnon2008mediate} noted that the baseline predictor could also be longitudinal. Recently, \citet{Liu2022mediate} introduced parallel bilinear growth models to examine mediation in nonlinear processes, employing a linear-linear piecewise function with an unknown knot to represent short- and long-term growth rates. These longitudinal mediation models assess each univariate process by estimating the regression coefficients of unidirectional paths between cross-outcome growth factors, unlike MGMs, which focus on covariances (correlations) between them.

Table \ref{tbl:Med_summary} outlines the specifications for two parallel linear growth models and two parallel bilinear growth models, including all possible paths within individual measurement occasions. Since the knot measurement conveys time-dependent effects, longitudinal mediation models with linear-linear functional forms use a reparameterization technique from \citet[Chapter~11]{Grimm2016growth} to unify pre- and post-knot expressions. Specifically, the initial status and two segment-specific slopes are reparameterized as the minimum value between $0$ and $t_{ij}-\gamma$, the measurement at the knot, and the maximum value between $0$ and $t_{ij}-\gamma$. This approach enables the estimation of both direct and indirect effects, as well as the total effect, providing a nuanced understanding of the relationships between predictors, mediators, and outcome variables over time.

\tablehere{5}

\subsubsection{Multiple-group Models for Longitudinal Processes}\label{spec:Multi}
Subsections \ref{spec:LGCMs} and \ref{spec:LCSMs} introduce LGCMs and LCSMs with TICs. When TICs are categorical, these models enable evaluation of differences in average growth trajectories between groups but may not fully capture variability in growth curves or differences in multivariate processes across such groups. This section presents the multiple-group modeling framework as an alternative for assessing group differences in various aspects of longitudinal processes, providing deeper insights into individual developmental differences.

With a specified model for each group, a one-group model can be easily extended to a multiple-group framework. Any model specified in Subsections \ref{spec:LGCMs}-\ref{spec:MED} can serve as a group-specific model, provided that the change pattern and model type remain consistent across groups. A general expression for a multiple-group model with $G$ groups is given by
\begin{equation}\label{eq:multigrp}
p(\text{sub-model}|mc_{i}=g)=\sum_{g=1}^{G}p(mc_{i}=g)\times p(\text{sub-model}|mc_{i}=g),
\end{equation}
where $mc_{i}$ represents the manifest class label for the $i^{th}$ individual, and $p()$ denotes the proportion of each group, summing to $1$ across groups.

\subsubsection{Mixture Models for Longitudinal Processes}\label{spec:GMMs}
Mixture models for longitudinal processes, often referred to as growth mixture models (GMMs) within the SEM framework, are similar to multiple-group models in that both relax the assumption of a single homogeneous population by capturing heterogeneity across subpopulations. However, unlike multiple-group models, where group membership is observed, GMMs treat group membership as latent, combining individuals from $G$ heterogeneous latent subpopulations. GMMs can be viewed as a clustering technique for longitudinal data, but unlike traditional methods like K-means, GMMs use a probability-based approach. Individuals are assigned to multiple latent classes with varying probabilities, with final classification based on the highest probability. These probabilities, which follow a multinomial distribution, can be regressed on TICs to inform group membership. In scenarios without such TICs, a general GMM with $G$ latent classes is specified as
\begin{equation}\label{eq:GMM1}
p(\text{sub-model}|lc_{i}=g)=\sum_{g=1}^{G}\pi(lc_{i}=g)\times p(\text{sub-model}|lc_{i}=g),
\end{equation}
where $lc_{i}$ is the latent class label for the $i^{th}$ individual, and $\pi()$ is the latent mixing component dividing the sample into $G$ classes, satisfying $0\le\pi()\le1$ and $\sum_{g=1}^{G}\pi()=1$.

For scenarios with TICs informing class formation, the specification becomes
\begin{align}
&p(\text{sub-model}|lc_{i}=g,\boldsymbol{x}_{gi})=\sum_{g=1}^{G}\pi(lc_{i}=g|\boldsymbol{x}_{gi})\times p(\text{sub-model}|lc_{i}=g),\label{eq:GMM2_1}\\
&\pi(lc_{i}=g|\boldsymbol{x}_{gi})=\begin{cases}
\frac{1}{1+\sum_{g=2}^{G}\exp(\beta_{g0}^{(g)}+\boldsymbol{\beta}_{g}^{(g)T}\boldsymbol{x}_{gi})}& \text{reference Group ($g=1$)}\\
\frac{\exp(\beta_{g0}^{(g)}+\boldsymbol{\beta}_{g}^{(g)T}\boldsymbol{x}_{gi})} {1+\sum_{g=2}^{G}\exp(\beta_{g0}^{(g)}+\boldsymbol{\beta}_{g}^{(g)T}\boldsymbol{x}_{gi})} & \text{other Groups ($g=2,\dots, G$)}
\end{cases},\label{eq:GMM2_2}
\end{align}
where $\boldsymbol{x}_{gi}$ is the TIC vector of the $i^{th}$ individual, and $\beta^{(g)}_{g0}$ and $\boldsymbol{\beta}^{(g)}_{g}$ are the intercept and coefficients of the multinomial logistic functions in the $g^{th}$ latent class.

Earlier works, such as \citet{Muthen1999GMM, Bouveyron2019GMM, Bauer2003GMM, Muthen2004GMM, Grimm2009FMM, Grimm2010FMM, Liu2019BLSGMM, Liu2021MoE, Liu2022PBLSGMM, Liu2023GMMTVC}, have developed GMMs with various sub-models, assessing their strengths and limitations. Most GMMs focus on heterogeneity in univariate longitudinal processes, though recent work has extended these models to multivariate longitudinal processes \citep{Liu2022PBLSGMM, Liu2023GMMTVC}. Building on these advancements, GMMs can incorporate any of the class-specific models discussed in Subsections \ref{spec:LGCMs}-\ref{spec:MED}, enabling the examination of heterogeneity in both univariate and multivariate longitudinal processes.

\subsection{Model Estimation}\label{sec:est}
All models specified in Section \ref{M:spec} require iterative model estimation, which relies on initial parameter values and involves calculating the expected model-implied mean vector $\boldsymbol{\mu}_{i}$ and variance-covariance structure $\boldsymbol{\Sigma}_{i}$. Model fit is assessed by comparing these to the observed data, using their difference as the function value. Maximum likelihood estimation (MLE), commonly known as full information maximum likelihood (FIML) in SEM research, is employed for this purpose. FIML calculates individual likelihoods and accommodates heterogeneity and missing data effectively.

The FIML function value is $-2\ln{L}$, where $L$ is the likelihood function across all $N$ individuals, with parameter space denoted as $\Theta$. For single-group analyses, assuming there are $p$ variables, where $p$ represents the total count of growth factors and the TICs that help explain the variability of these growth factors, measured for the $i^{th}$ individual in a vector $\boldsymbol{\omega}_{i}$, $L$ is defined as follows, $L$ is defined as
\begin{equation}\label{eq:lik1}
L(\Theta)=\prod^{N}_{i=1}\frac{\exp{\big(-\frac{1}{2}(\boldsymbol{\omega}_{i}-\boldsymbol{\mu}_{i})^{T}\boldsymbol{\Sigma}^{-1}_{i}(\boldsymbol{\omega}_{i}-\boldsymbol{\mu}_{i})\big)}}{\sqrt{(2\pi)^{p}|\boldsymbol{\Sigma}_{i}|}},
\end{equation}
where $\boldsymbol{\mu}_{i}$ and $\boldsymbol{\Sigma}_{i}$ are the model-implied mean vector and variance-covariance structure for individual $i$. 

For a multiple-group model with $G$ manifested groups, the likelihood function across all individuals is defined as
\begin{equation}\label{eq:lik2}
L(\Theta)=\prod^{N}_{i=1}\bigg(\sum^{G}_{g=1}p(mc_{i}=g)\times\frac{\exp{\big(-\frac{1}{2}(\boldsymbol{\omega}_{i}-\boldsymbol{\mu}^{(g)}_{i})^{T}\boldsymbol{\Sigma}^{(g)-1}_{i}(\boldsymbol{\omega}_{i}-\boldsymbol{\mu}^{(g)}_{i})\big)}}{\sqrt{(2\pi)^{p}|\boldsymbol{\Sigma}^{(g)}_{i}|}}\bigg),
\end{equation}
where $\boldsymbol{\mu}^{(g)}_{i}$ and $\boldsymbol{\Sigma}^{(g)}_{i}$ are the mean vector and variance-covariance structure for individual $i$ in group $g$. For a growth mixture model with $G$ latent classes, the likelihood function across all $N$ individuals is defined as
\begin{equation}\label{eq:lik3}
L(\Theta)=\prod^{N}_{i=1}\bigg(\sum^{G}_{g=1}\pi(lc_{i}=g)\times\frac{\exp{\big(-\frac{1}{2}(\boldsymbol{\omega}_{i}-\boldsymbol{\mu}^{(g)}_{i})^{T}\boldsymbol{\Sigma}^{(g)-1}_{i}(\boldsymbol{\omega}_{i}-\boldsymbol{\mu}^{(g)}_{i})\big)}}{\sqrt{(2\pi)^{p}|\boldsymbol{\Sigma}^{(g)}_{i}|}}\bigg)
\end{equation}
and 
\begin{equation}\label{eq:lik4}
L(\Theta)=\prod^{N}_{i=1}\bigg(\sum^{G}_{g=1}\pi(lc_{i}=g|\boldsymbol{x}_{gi})\times\frac{\exp{\big(-\frac{1}{2}(\boldsymbol{\omega}_{i}-\boldsymbol{\mu}^{(g)}_{i})^{T}\boldsymbol{\Sigma}^{(g)-1}_{i}(\boldsymbol{\omega}_{i}-\boldsymbol{\mu}^{(g)}_{i})\big)}}{\sqrt{(2\pi)^{p}|\boldsymbol{\Sigma}^{(g)}_{i}|}}\bigg)
\end{equation}
for scenarios without and with the TICs informing cluster formation (i.e., $\boldsymbol{x}_{gi}$), in which $\boldsymbol{\mu}^{(g)}_{i}$ and $\boldsymbol{\Sigma}^{(g)}_{i}$ are the submodel-implied mean vector and variance-covariance structure of the $i^{th}$ individual in the $g^{th}$ latent class. 

\section{Package Implementation}\label{sec:imp}
This section describes how to implemented the models specified earlier within the package. It covers the estimation functions for all models specified in Section \ref{M:spec}, detailing how they are initialized and optimized. Additionally, we discuss the methods for deriving non-estimable parameters and the post-fit computation functions.

\subsection{Estimation Functions}\label{I:est}
This package currently provides seven estimation functions for the models specified in Section \ref{M:spec}: 
\begin{enumerate}
\item {\code{getLGCM()} for latent growth curve models (LGCMs), without or with time-invariant covariates (TICs),}
\item {\code{getLCSM()} for latent change score models (LCSMs), without or with TICs,}
\item {\code{getTVCmodel()} for LGCMs or LCSMs with a time-varying covariate (TVC),}
\item {\code{getMGM()} for multivariate version of LGCMs or LCSMs,}
\item {\code{getMediation()} for longitudinal mediation analysis,}
\item {\code{getMGroup()} for multiple-group version of the models (1)-(5), and}
\item {\code{getMIX()} for mixture model version of the models (1)-(5).}
\end{enumerate}

In general, the arguments in these estimation functions include: (1) a dataset in wide format, (2) longitudinal variables and their corresponding time variables, (3) covariates, including TICs and TVCs, (4) information on manifested or latent groups when applicable, and (5) settings for initializing parameters and optimizing the model (e.g., initial values, optimization status, and retry mechanisms, see details in Section \ref{sec:opt}). Detailed information on the syntax, arguments, and usage of each function is available in the package documentation. Users are encouraged to refer to the package help files and vignettes for comprehensive guidance on implementing specific models.

In addition, users interested in the underlying \pkg{OpenMx} specifications can directly access the source code for internal functions in \pkg{nlpsem}. For example, \code{nlpsem:::getUNI.loadings} and \code{nlpsem:::getMULTI.loadings} allow users to view how factor loadings are specified with definition variables for univariate and multivariate longitudinal models, respectively. In addition, \href{https://github.com/Veronica0206/nlpsem_manuscript/tree/main/Demo_for_OpenMx}{\pkg{OpenMx}} and \href{https://github.com/Veronica0206/nlpsem_manuscript/tree/main/Demo_for_Mplus8}{\pkg{Mplus 8}} code for selected models are available online for researchers interested in exploring or benchmarking these implementations further.

\subsection{Model Initialization and Optimization}\label{sec:opt}
Iterative estimation algorithms require initializing the parameter space $\Theta$ with suitable initial values, which improves the likelihood of convergence and reduces computational load. In \pkg{nlpsem}, the \code{starts} argument controls the initialization process, allowing users to specify initial values or derive them from raw data. For single-group models, initial growth factors are derived by fitting linear regressions (\code{lm()}) for linear functions or nonlinear regressions (\code{nls()}) for nonlinear functions on individual data.

The initialization process depends on the model specification. For models without covariates or mediators, the mean values and variance-covariance matrices of the `raw' growth factors are calculated. In models with time-invariant covariates (TICs), path coefficients are initialized by regressing the `raw' growth factors on TICs. When the model includes a time-varying covariate (TVC), longitudinal outcomes are regressed on the TVC measurements, while baseline and temporal effects are separately initialized for decomposed TVCs. For longitudinal mediation models, initialization is carried out in stages: first, the mediator’s growth factors are regressed on predictor-related parameters, followed by regressing the outcome’s growth factors on both predictor and mediator-related parameters. In multiple-group models, these steps are repeated for each manifest group. In mixture models, the initialization process begins with the K-means algorithm to assign individuals to latent classes, after which the aforementioned steps are applied within each class. Residual variances and covariances, which cannot be directly derived from raw data, require user input through the \code{res_scale} and \code{res_cor} arguments.

Once initialization is complete, optimization is performed iteratively using the built-in engines of \pkg{OpenMx}, which minimize the FIML function value until convergence is achieved, as indicated by a status code of $0$ \citep{OpenMx2016package, Pritikin2015OpenMx, Hunter2018OpenMx, User2020OpenMx}. The \pkg{OpenMx} framework provides three optimization engines: \code{NPSOL}, a robust optimizer developed by the Nonlinear Programming group at the Systems Optimization Laboratory \citep{Gill1986npsol}; \code{SLSQP}, which is based on Sequential Least-Squares Quadratic Programming \citep{Johnson2014nlopt, Kraft1994}; and \code{CSOLNP}, a C++-based optimizer for SOLving Non-linear Programs \citep{Zahery2017csolnp}. Among these, \code{CSOLNP} is recommended for its efficiency in handling non-linearly constrained problems, while \code{SLSQP} is preferred for generating likelihood-based confidence intervals due to its robustness in constrained optimization scenarios. 

\subsection{Additional Non-estimable Parameters}\label{sec:non-est}
The \pkg{nlpsem} package enables the estimation of non-estimable parameters by leveraging the \code{mxAlgebra()} and \code{mxSE()} functions from the \pkg{OpenMx} package. The \code{mxAlgebra()} function calculates point estimates for parameters derived from free parameters, while \code{mxSE()} computes corresponding standard errors using the (multivariate) delta method \citep{OpenMx2016package, Pritikin2015OpenMx, Hunter2018OpenMx, User2020OpenMx}. The package supports four categories of non-estimable parameters:

\begin{enumerate}
\item \textbf{Reparameterized Parameters}: In bilinear spline growth curve models with an unknown knot (as discussed in Subsection \ref{spec:LGCMs}), reparameterization unifies pre- and post-knot expressions, though these reparameterized parameters may lack direct interpretability. The package \pkg{nlpsem} integrates inverse transformation functions and matrices from \citet{Liu2019BLSGM} and \citet{Liu2021PBLSGM} to restore these parameters to their original, interpretable forms.
\item \textbf{Interval-Specific Parameters}: As introduced in Subsection \ref{spec:LCSMs}, LCSMs allow for the estimation of means and variances of interval-specific slopes, interval-specific changes, and change-from-baseline values. These can be treated as additional parameters.
\item \textbf{Conditional Distributions}: In models with covariates (or a mediator), such as LGCMs with a TIC, the package estimates growth factor intercepts and unexplained variance-covariance structures (as shown in Equation \ref{eq:LGCM2_TIC}). However, the conditional mean vector of growth factors, which directly relates to developmental theory, is often of greater interest. The package facilitates the derivation of these conditional parameters.
\item \textbf{Indirect and Total Effects}: The package also estimates indirect and total effects of predictors on outcomes in longitudinal mediation models.
\end{enumerate}

\subsection{Post-fit Computations}\label{sec:post}
The \pkg{nlpsem} package provides a set of post-fit computations, analyses, and evaluations to complement its estimation functions. These tools are designed to enhance the interpretability and utility of model results. Specific analyses, such as posterior classification, enumeration processes, and latent kappa statistics \citep{Dumenci2011kappa, Dumenci2019knee}, are particularly tailored for mixture models. This section provides an overview of key post-fit computations. Detailed arguments and usage instructions for all post-fit computation functions are provided in the package documentation.

\subsubsection{Statistical Significance: Wald p-values and Confidence Intervals}\label{fit:sig}
Quantifying the uncertainty of estimates is crucial in statistical analysis. The \pkg{nlpsem} package generates point estimates and standard errors for all estimable or derived parameters by default. The \code{getEstimateStats()} function calculates Wald p-values \citep{Wald1943wald} and supports three types of confidence intervals: Wald \citep{Casella2002wald}, likelihood-based (or `likelihood profile') \citep{Madansky1965likelihood, Matthews1988likelihood}, and bootstrap intervals \citep[Chapter~12]{Efron1994bootstrap}. While Wald intervals assume asymptotic normality, likelihood-based and bootstrap intervals relax this assumption, ensuring robustness for small sample sizes or non-normal distributions. These computations utilize \pkg{OpenMx} tools such as \code{mxCI()} and \code{mxBootstrap()} \citep{OpenMx2016package, Pritikin2015OpenMx, Hunter2018OpenMx, User2020OpenMx}.

\subsubsection{Model Selection between Intrinsically Nonlinear Longitudinal Models and Their Parsimonious Alternatives}\label{fit:LRT}
Fitting intrinsically nonlinear longitudinal models often requires balancing complexity with interpretability. Parsimonious alternatives, such as models with fixed growth rate ratio, growth acceleration ratio, or knot, are typically nested within their intrinsically nonlinear counterparts. The likelihood ratio test (LRT) is a core method for comparing these nested models, allowing researchers to evaluate whether the simpler, nested model sufficiently addresses the research objectives while maintaining statistical robustness. In addition to the regular LRT, the \code{getLRT()} function, built on \pkg{OpenMx}'s \code{mxCompare()}, also supports bootstrap LRTs, providing more accurate p-values when the standard assumptions of the LRT are not fully met \citep{Feng1996bootstrap}.

\subsubsection{Derivation of Individual Factor Scores}\label{fit:FSest}
In longitudinal data analysis, between-individual differences are captured by the variance of growth factors. The \code{getIndFS()} function, built on \pkg{OpenMx}'s \code{mxFactorScores()} \citep{OpenMx2016package, Pritikin2015OpenMx, Hunter2018OpenMx, User2020OpenMx}, estimates factor scores for all specified latent variables. These include free latent variables (e.g., growth factors) and derived latent variables (e.g., interval-specific slopes and changes from baseline), along with their standard errors.

\subsubsection{Posterior Classification}\label{fit:post}
Posterior classification is critical for evaluating latent class membership in mixture models. Using Bayes' rule, the posterior probabilities for each latent class are calculated based on either:
\begin{equation}\label{eq:post1}
p(lc_{i}=g)=\frac{\pi(lc_{i}=g)\times p(\text{sub-model}|lc_{i}=g)}{\sum_{g=1}^{G}\pi(lc_{i}=g)\times p(\text{sub-model}|lc_{i}=g)}.
\end{equation}
or, if TICs are included:
\begin{equation}\label{eq:post2}
p(lc_{i}=g)=\frac{\pi(lc_{i}=g|\boldsymbol{x}_{gi})\times p(\text{sub-model}|lc_{i}=g)}{\sum_{g=1}^{G}\pi(lc_{i}=g|\boldsymbol{x}_{gi})\times p(\text{sub-model}|lc_{i}=g)}.
\end{equation}

\subsubsection{Model Selection and Enumeration Process with Summary Tables}\label{fit:summary}
The \code{getSummary()} function facilitates model comparison and selection using metrics like Akaike Information Criterion (AIC) and Bayesian Information Criterion (BIC). For mixture models, the enumeration process involves fitting a series of candidate models with varying numbers of latent classes and selecting the optimal model based on BIC \citep{Nylund2007BIC}. It is generally recommended to perform enumeration without covariates to avoid confounding effects \citep{Diallo2017TICs, Nylund2016TICs}.

\subsubsection{Visualization of Estimated Growth Curves and Growth Changes Over Time}
The \code{getFigure()} function enables visualization of estimated growth trajectories and changes over time, including class-specific estimates for multiple group models and mixture models. These plots provide clear insights into developmental patterns and facilitate interpretation of complex longitudinal data.

\subsubsection{Latent Kappa Statistic}\label{fit:kappa}
The latent kappa statistic \citep{Dumenci2011kappa} measures the consistency of latent class assignments across clustering algorithms or models. Its applications include evaluating the impact of TICs on latent class membership \citep{Liu2021MoE} and comparing cluster assignments derived from different longitudinal outcomes \citep{Dumenci2019knee, Liu2022PBLSGMM}. The \code{getLatentKappa()} function in \pkg{nlpsem} supports these analyses, providing robust tools for assessing individual heterogeneity across models.

\section{Tutorial: Synthetic Examples}\label{sec:example}
\subsection{Example Data}
The \pkg{nlpsem} package includes a sample dataset, \code{RMS_dat}, derived from the publicly accessible portion of the Early Childhood Longitudinal Study, Kindergarten Class of 2010-11 (ECLS-K:2011). Conducted by the National Center for Education Statistics (NCES), this study encompassed nine rounds of data collection from the fall of 2010 to the spring of 2016. The ECLS-K:2011 dataset provides a wealth of information about children's early life experiences, including health, developmental milestones, educational progress, and pre-kindergarten experiences\footnote{This dataset has been formatted solely for demonstration purposes and excludes the original survey weights. Researchers conducting comprehensive analyses should refer to the full dataset available on the NCES data products page and ensure proper application of complex survey weights.}.

The \code{RMS_dat} dataset comprises 500 observations across 49 variables, covering a wide range of factors relevant to early childhood education research. These variables include academic performance metrics, such as reading, mathematics, and science scores across study waves, as well as demographic and environmental characteristics like age-in-month at each wave, sex, race, and family income. Teacher evaluations of children's behavioral and learning traits, including approach to learning, self-control, interpersonal skills, and attention focus, are also included. To enhance computational and interpretative clarity, the dataset underwent three key preprocessing steps. First, the initial time point was set to zero for all time-varying measurements, allowing the initial status estimates to represent values at the onset of the study. This was implemented in R by adjusting each measurement occasion (\code{T1} through \code{T9}) relative to the initial time point (\code{T1}). Second, time-invariant covariates (TICs) were standardized using the  \code{scale()} R function, centering the variables at a mean of zero and scaling them to have a standard deviation of one, thereby expediting computations. Finally, the time-varying covariate (TVC), represented by reading scores (\code{R1} through \code{R9}), was standardized following the approach outlined by \citet{Liu2022decompose, Liu2023GMMTVC}. Specifically, the mean and variance of the baseline reading score (\code{R1}) were used to standardize the scores across all time points. The R code and corresponding output for these preprocessing steps are available in the package's online documentation hosted on \href{https://github.com/Veronica0206/nlpsem_manuscript/blob/main/Demo_for_nlpsem/Synthetic-Examples.md}{GitHub}.

\subsection{getLGCM() Examples}
The \code{getLGCM()} function applies latent growth curve models (LGCMs) to analyze mathematics development using linear-linear trajectories without covariates. In this section, we fit two models: an intrinsically nonlinear model (a linear-linear function with a random knot) and its parsimonious counterpart (a linear-linear function with a fixed knot). To compare the models, we performed a likelihood ratio test, evaluating whether the simpler model provides an adequate alternative to the more complex one without significantly compromising fit or explanatory power. Additionally, we visualized the estimated developmental trajectories of mathematics ability for both models (see Figures \ref{fig:LGCM1} and \ref{fig:LGCM2}). As shown in these figures, both LGCMs effectively capture the underlying patterns of mathematics development. However, comparative metrics such as AIC, BIC, and the likelihood ratio test suggest that the intrinsically nonlinear model provides a superior fit for modeling mathematics ability. The R code and corresponding output for this analysis are available on \href{https://github.com/Veronica0206/nlpsem_manuscript/blob/main/Demo_for_nlpsem/Synthetic-Examples.md}{GitHub}.

\figurehere{1}

\subsection{getLCSM() Examples}
The \code{getLCSM()} function demonstrates reading development using latent change score models (LCSMs) with a nonparametric functional form. In this section, we fit two models: one without covariates and another incorporating two standardized growth time-invariant covariates (TICs): baseline teacher-reported approach to learning and attentional focus. For both models, we provide a summary table and, for the model with growth TICs, compute p-values and confidence intervals. Additionally, we visualize the estimated change-from-baseline and growth rate for the model without TICs in Figure \ref{fig:LCSM}. The R code and corresponding output for these analyses are available on \href{https://github.com/Veronica0206/nlpsem_manuscript/blob/main/Demo_for_nlpsem/Synthetic-Examples.md}{GitHub}.

\figurehere{2}

Figures \ref{fig:LCSM1} and \ref{fig:LCSM2} illustrate the estimated values for both change-from-baseline and interval-specific growth rates, derived from the estimated growth factors. It is important to note that the growth factor estimates from the model incorporating growth TICs are conditional on these TICs and, therefore, differ from those of the model excluding TICs. A review of the summary table shows that Model 1 (the nonparametric LCSM without growth TICs) has smaller estimated likelihood, AIC, and BIC values compared to Model 2 (the nonparametric LCSM with two growth TICs). However, these differences do not imply that Model 1 outperforms Model 2, as the data used for fitting the two models differ. By setting \code{CI_type = "all"} in the \code{getEstimateStats()} function, three types of confidence intervals—Wald, likelihood-based, and bootstrap—are generated. It is worth noting that likelihood-based and bootstrap confidence intervals are computed only for free parameters, whereas Wald confidence intervals are provided for both free parameters and those that are non-estimable. These non-estimable parameters include the conditional mean values of the growth factors, as well as the means and variances of interval-specific slopes, interval-specific changes, and changes from baseline.

\subsection{getTVCmodel() Examples}
The function \code{getTVCmodel()} is designed to construct univariate longitudinal outcome models that incorporate a time-varying covariate (TVC). It exemplifies a latent growth curve model with an intrinsically linear-linear functional form for mathematics development while simultaneously evaluating the influence of reading ability on mathematics development over time. This section includes the fitting of two distinct models: both treat reading ability as the TVC and the baseline teacher-reported approach to learning as the time-invariant covariate (TIC). However, the first model directly regresses on the TVC, while the second decomposes the TVC into its baseline value and a set of interval-specific slopes. In addition to fitting these models, we computed p-values and Wald confidence intervals for the model that incorporates the decomposed TVC. Figures \ref{fig:TVC1} and \ref{fig:TVC2} present the estimated growth trajectories of mathematics development for both models. The R code and corresponding output are available on \href{https://github.com/Veronica0206/nlpsem_manuscript/blob/main/Demo_for_nlpsem/Synthetic-Examples.md}{GitHub}.

\figurehere{3}

The results highlight the advantages of incorporating a decomposed time-varying covariate (TVC) into the analysis. First, the decomposed TVC allows for the examination of both baseline and temporal effects of reading ability on mathematics development. According to the model results, the baseline effect of reading ability on early-stage mathematics development is estimated at $0.0708$, indicating that for every standardized unit increase in baseline reading ability, the growth rate of mathematics ability before Grade 3 increases by $0.0708$. The temporal effect of reading ability on mathematics development is estimated at $21.3467$, suggesting that the final mathematics score in the spring semester of Grade 1 increases by $21.3467$ for each standardized unit growth in the reading ability growth rate within that semester.

Second, the decomposed model facilitates an exploration of the relationship between the TVC baseline value and the TIC. For instance, the covariance between baseline reading ability and the teacher-reported approach to learning was estimated at $0.4000$ at the beginning of the ECLS-K:2011 study. Notably, as shown in Figure \ref{fig:TVC}, incorporating a TVC into a longitudinal model tends to underestimate growth factors and trajectories since the longitudinal outcome is regressed on the TVC (or its temporal states). However, the extent of underestimation is significantly reduced in the model with a decomposed TVC (Figure \ref{fig:TVC2}) compared to the model that directly integrates the TVC (Figure \ref{fig:TVC1}).

\subsection{getMGM() Examples}
The \code{getMGM()} function constructs a multivariate growth model (MGM) to analyze the development of reading and mathematics abilities, as well as the correlations between these two developmental processes over time. In addition to fitting the model, we computed p-values and Wald confidence intervals for this bivariate longitudinal model. Figures \ref{fig:MGM1} and \ref{fig:MGM2} display visualizations of the estimated growth trajectories for both reading and mathematics abilities. Beyond the growth factors associated with each univariate developmental process, this model estimates the covariances between the growth factors of different outcomes and the residual covariance. The R code and corresponding output are available on \href{https://github.com/Veronica0206/nlpsem_manuscript/blob/main/Demo_for_nlpsem/Synthetic-Examples.md}{GitHub}. The model output highlights positive relationships between the developmental processes of reading and mathematics abilities. This is evidenced by positive intercept-intercept (YZ\_psi00, $\text{p-value}<0.0001$) and pre-knot slope-slope (YZ\_psi11, $\text{p-value}<0.0001$) covariances. These findings indicate that students who demonstrated higher reading ability at the start of the ECLS-K:2011 study also tended to exhibit higher mathematics ability, and vice versa. Similarly, students with more rapid growth in reading ability during the early stages were generally associated with more rapid growth in mathematics ability, and vice versa.

\figurehere{4}

\subsection{getMediation() Examples}
The \code{getMediation()} function constructs longitudinal mediation models. In this section, we develop two such models. The first model, utilizing a linear-linear functional form and a baseline predictor, examines how the baseline approach to learning influences mathematics development through the mediation of reading ability development. The second model, also adopting a linear-linear functional form but with a longitudinal predictor, investigates how the developmental trajectory of reading ability impacts the development of science ability through the mediation of mathematics development. Both models include p-values and Wald confidence intervals to substantiate the findings. The R code and corresponding output are available on \href{https://github.com/Veronica0206/nlpsem_manuscript/blob/main/Demo_for_nlpsem/Synthetic-Examples.md}{GitHub}.

Similar to MGMs, these longitudinal mediation models estimate growth trajectories for each univariate developmental process and the relationships among these processes over time. However, the relationships between the processes are captured through the coefficients of unidirectional paths. For example, the output from the first model indicates that the baseline approach to learning positively influences the early-stage growth rate of reading ability (betaM1, $\text{p-value}=0.0070$), as well as reading ability at the knot (betaMr, $\text{p-value}<0.0001$) and mathematics ability at the knot (betaYr, $\text{p-value}=0.0119$). Additionally, the early-stage growth rate of reading ability positively impacts the early-stage growth rate of mathematics ability (betaM1Y1, $\text{p-value}<0.0001$).

With these path coefficients, we calculated both the indirect effect (mediation effect) of the baseline approach to learning on mathematics development through reading ability development and the total effect of the approach to learning on mathematics development. For instance, through the early growth rate of reading ability, the indirect effect of the baseline approach to learning on the early growth rate of mathematics ability is $0.0237$. Consequently, the total effect of the baseline approach to learning on the early growth rate of mathematics ability is $0.0386$ ($0.0386=0.0149+0.0237$). The path coefficients, indirect effects, and total effects for the second longitudinal model can be interpreted similarly.

\subsection{getMGroup() Examples}
The \code{getMGroup()} function constructs a multiple-group latent growth curve model to examine group differences in developmental trajectories. This model utilizes a linear-linear functional form with a random knot to analyze differences in mathematics development from Grade K through Grade 5 between boys and girls. The R code and corresponding output are available on \href{https://github.com/Veronica0206/nlpsem_manuscript/blob/main/Demo_for_nlpsem/Synthetic-Examples.md}{GitHub}. As illustrated in Figure \ref{fig:MGroup}, boys demonstrate a slightly faster development in mathematical ability compared to girls. However, this difference is not statistically significant, as indicated by the overlapping confidence intervals between the two manifest groups.

\figurehere{5}

\subsection{getMIX() Examples}
The \code{getMIX()} function is used to construct mixture latent growth curve models with a linear-linear functional form and a fixed knot. As part of the analysis, we conducted an enumeration process, fitting models with one to three latent classes. By utilizing the \code{getSummary()} function with \code{HetModels = TRUE}, we obtained estimates of the likelihood, AIC, BIC, class-specific residuals, and class-specific proportions for each model. Both the likelihood estimates and information criteria consistently identified the three-class model as optimal.

The estimated growth trajectories for each of the three latent classes are depicted in Figure \ref{fig:MIX}. The R code and corresponding output are available on \href{https://github.com/Veronica0206/nlpsem_manuscript/blob/main/Demo_for_nlpsem/Synthetic-Examples.md}{GitHub}. The diagram reveals that students in the third latent class outperformed the other two classes in mathematics. Although the other two classes showed overlapping patterns in mathematics development during the early stage, the growth pace of the first group slowed down earlier than the second group. 

\section{Concluding Remarks}\label{sec:conclude}
The developed R package, \pkg{nlpsem}, aims to facilitate comprehensive evaluations of nonlinear longitudinal processes, including intrinsically nonlinear functional forms within the SEM framework. It currently supports three commonly used intrinsically nonlinear functional forms, namely, the individual ratio of growth rate under the negative exponential function, the individual ratio of growth acceleration within the Jenss-Bayley function, and the individual knot in the bilinear spline function, also known as the linear-linear piecewise function. In addition, this package is versatile enough to handle parsimonious models and models with a quadratic functional form, which belong to Type II and Type I of nonlinear longitudinal models, respectively. Despite not primarily focusing on models for linear longitudinal processes, \pkg{nlpsem} incorporates functionalities for them, making it a comprehensive tool for researchers in the field. The package provides computational resources for univariate longitudinal processes, with the option to include or exclude time-invariant covariates. Further, it facilitates estimations for multivariate longitudinal processes, including a longitudinal outcome with time-varying covariates, correlated growth models for multiple outcomes, and longitudinal mediation models. Multiple group and mixture models are accommodated within \pkg{nlpsem}, where the sub-model can be any of the types above. Built on the \pkg{OpenMx} package, it enables flexible SEM specification and data-driven parameter estimation through built-in optimizers. Note that the package allows for unstructured time frame compatibility by employing the definition variables approach.

Despite its capabilities, \pkg{nlpsem} has limitations, which also pave the way for future developments. First, other nonlinear functional forms, such as logistic and Gompertz functions, are not currently supported. The inclusion of such additional forms could enhance the flexibility and applicability of the package. Second, formal statistical hypothesis testing needs to be developed for complex longitudinal models to evaluate the impact of removing or adding specific paths on the overall model. Third, although several nonlinear longitudinal models with certain functional forms have been well-documented and validated by simulation studies, others still need to be explored. Conducting further simulation studies to investigate the performance of these models under different scenarios will provide essential insights and improve the robustness and validity of the \pkg{nlpsem} package.

\bibliography{Extension11}

\renewcommand\thetable{\arabic{figure}}
\setcounter{figure}{0}

\begin{figure}[!htbp]
\centering
    \subfloat[Bilinear Growth Curve (Random Knot)]{\includegraphics[width=0.5\textwidth]{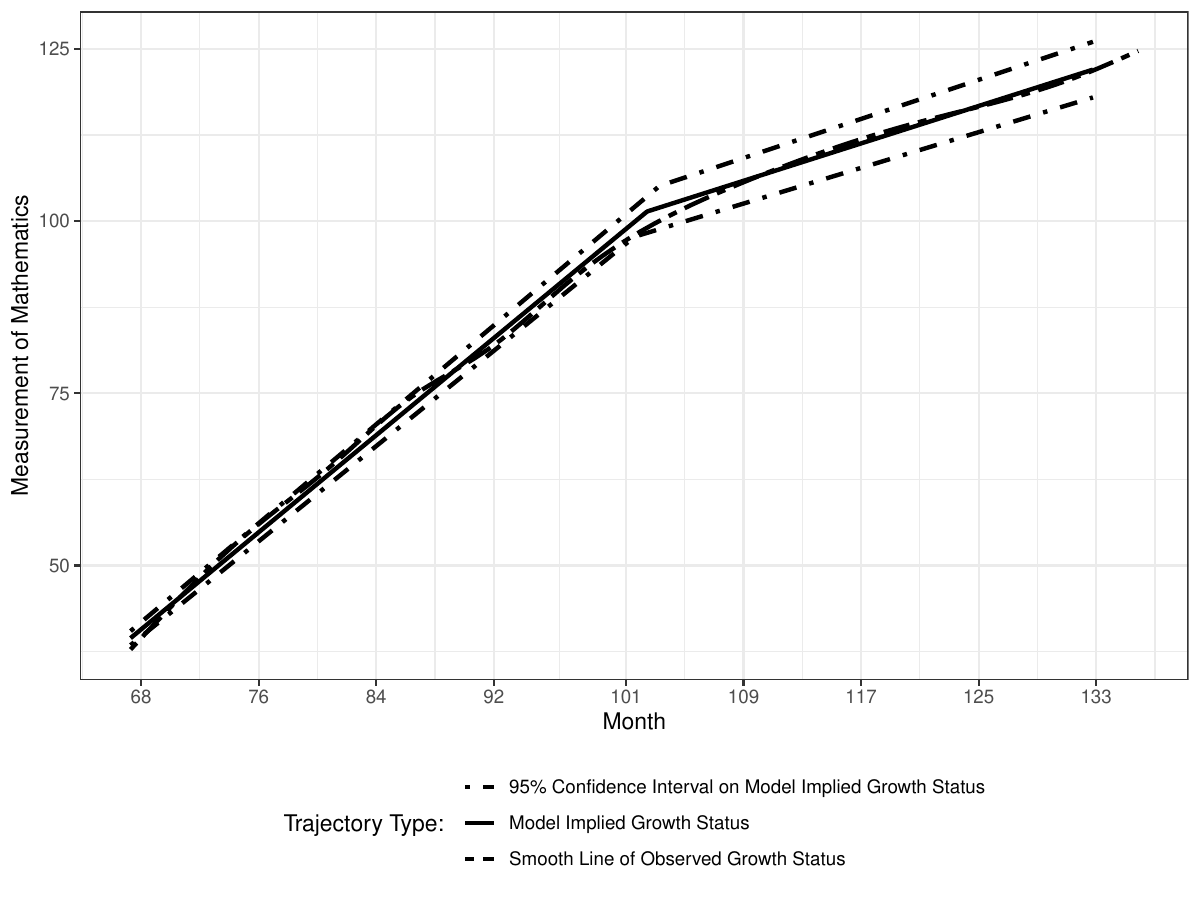}\label{fig:LGCM1}}
\hfill
    \subfloat[Bilinear Growth Curve (Fix Knot)]{\includegraphics[width=0.5\textwidth]{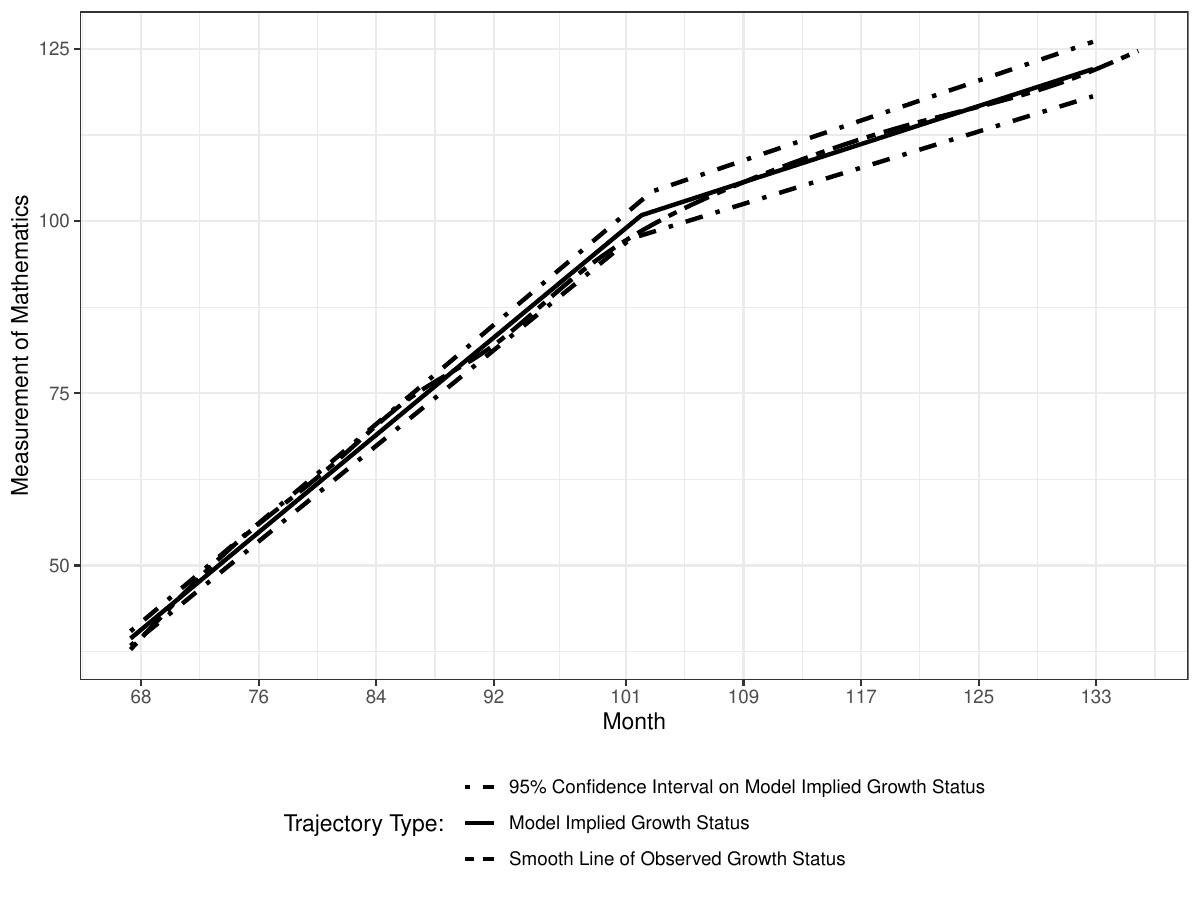}\label{fig:LGCM2}}
\caption{Estimated Development Status of Mathematics Ability from Latent Growth Curve Models}
\label{fig:LGCM}
\end{figure}

\begin{figure}[!htbp]
\centering
    \subfloat[Estimated Change from Baseline]{\includegraphics[width=0.5\textwidth]{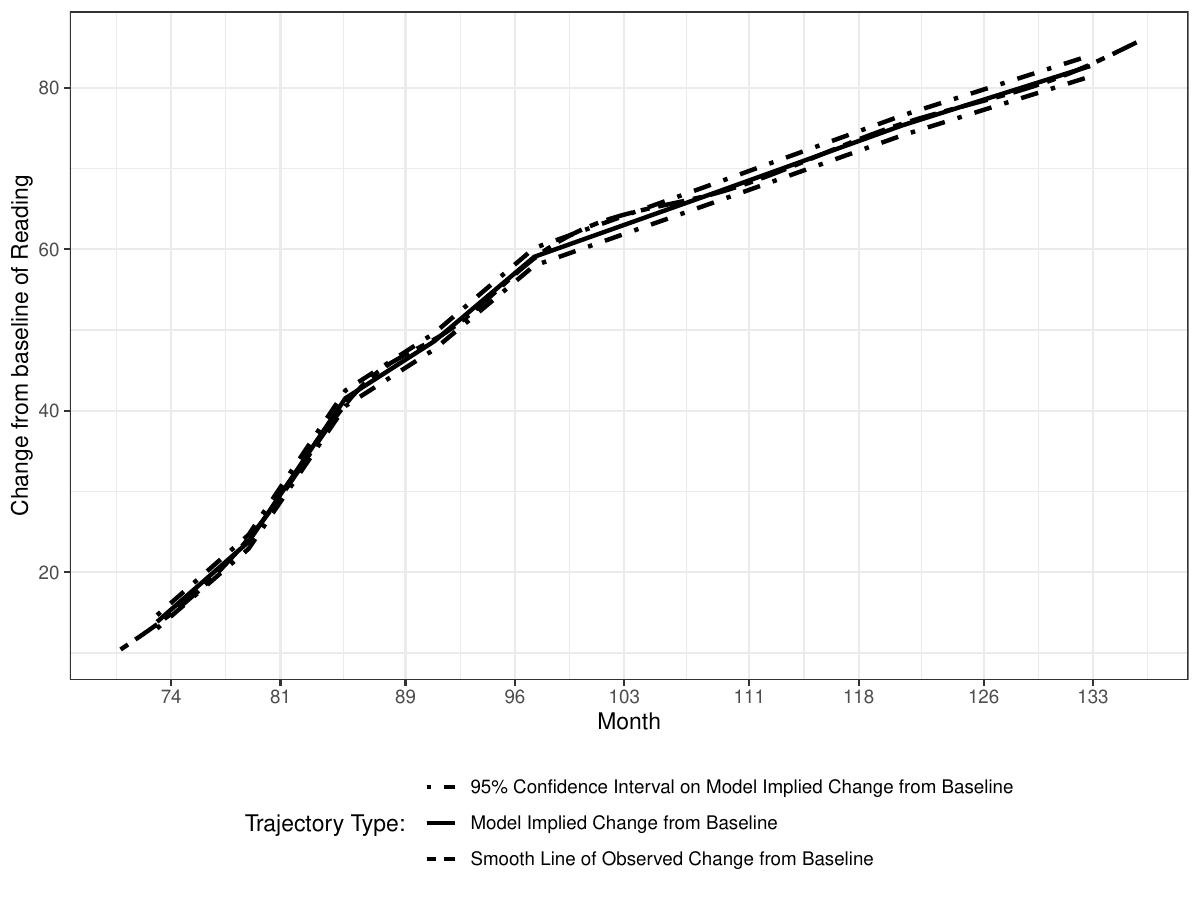}\label{fig:LCSM1}}
\hfill
    \subfloat[Estimated Growth Rate]{\includegraphics[width=0.5\textwidth]{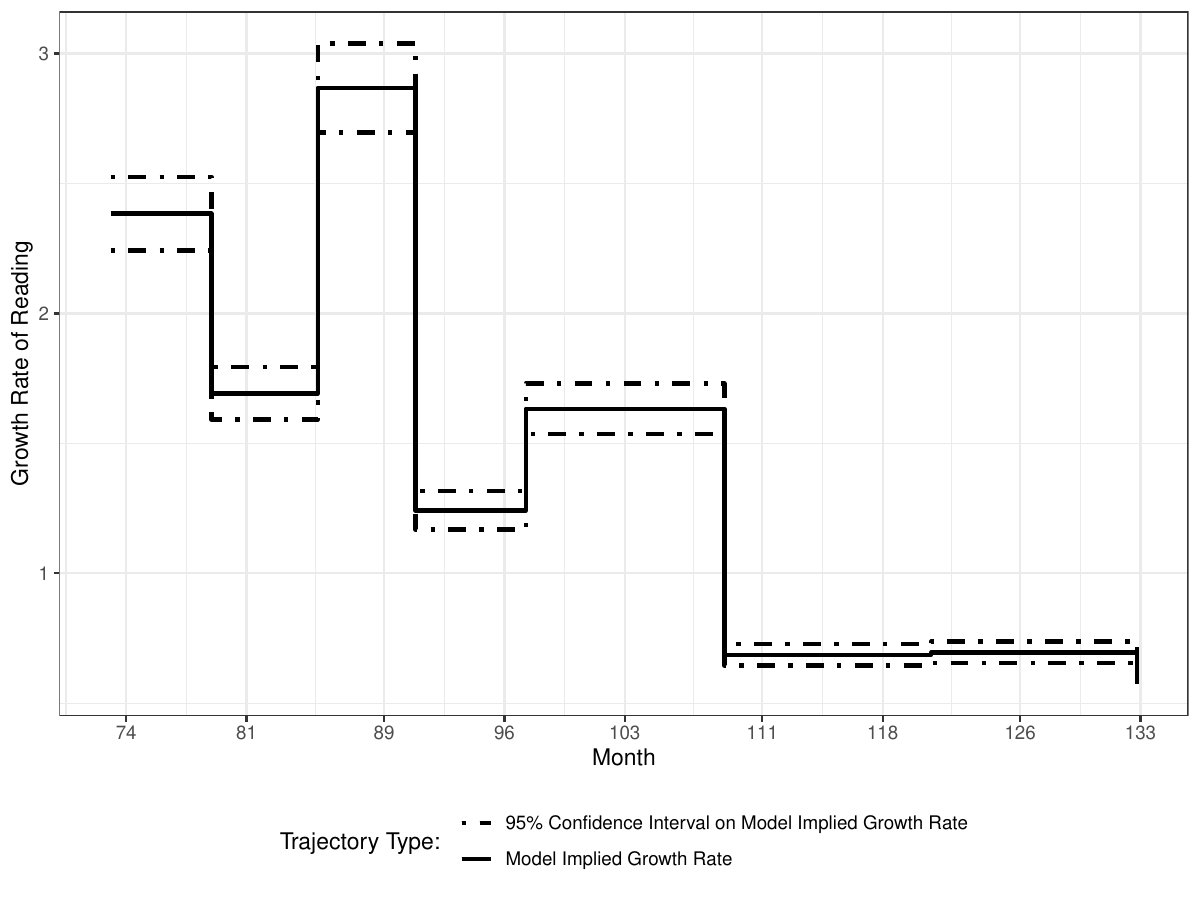}\label{fig:LCSM2}}
\caption{Latent Change Score Models with Nonparametric Function for Reading Ability}
\label{fig:LCSM}
\end{figure}

\begin{figure}[!htbp]
\centering
    \subfloat[Inclusion of a TVC]{\includegraphics[width=0.5\textwidth]{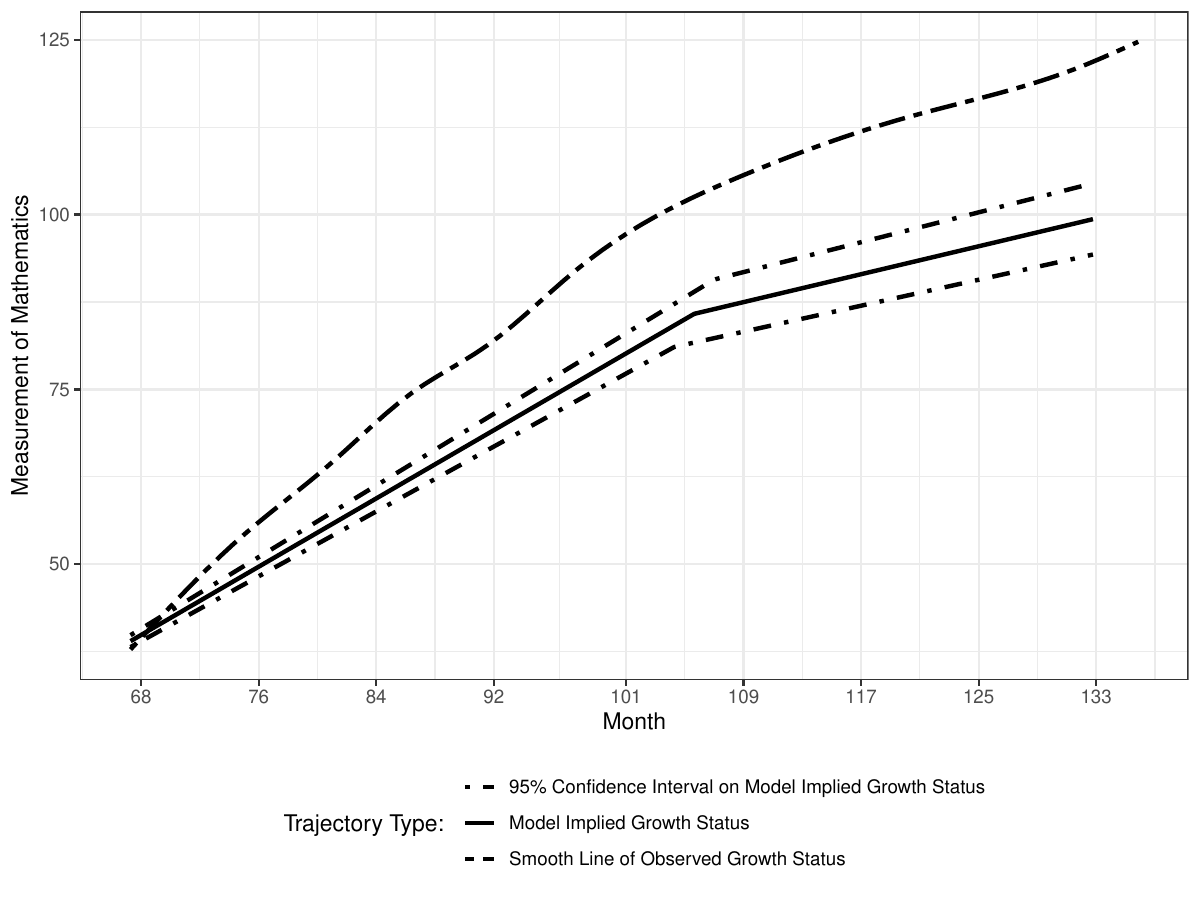}\label{fig:TVC1}}
\hfill
    \subfloat[Inclusion of a Decomposed TVC]{\includegraphics[width=0.5\textwidth]{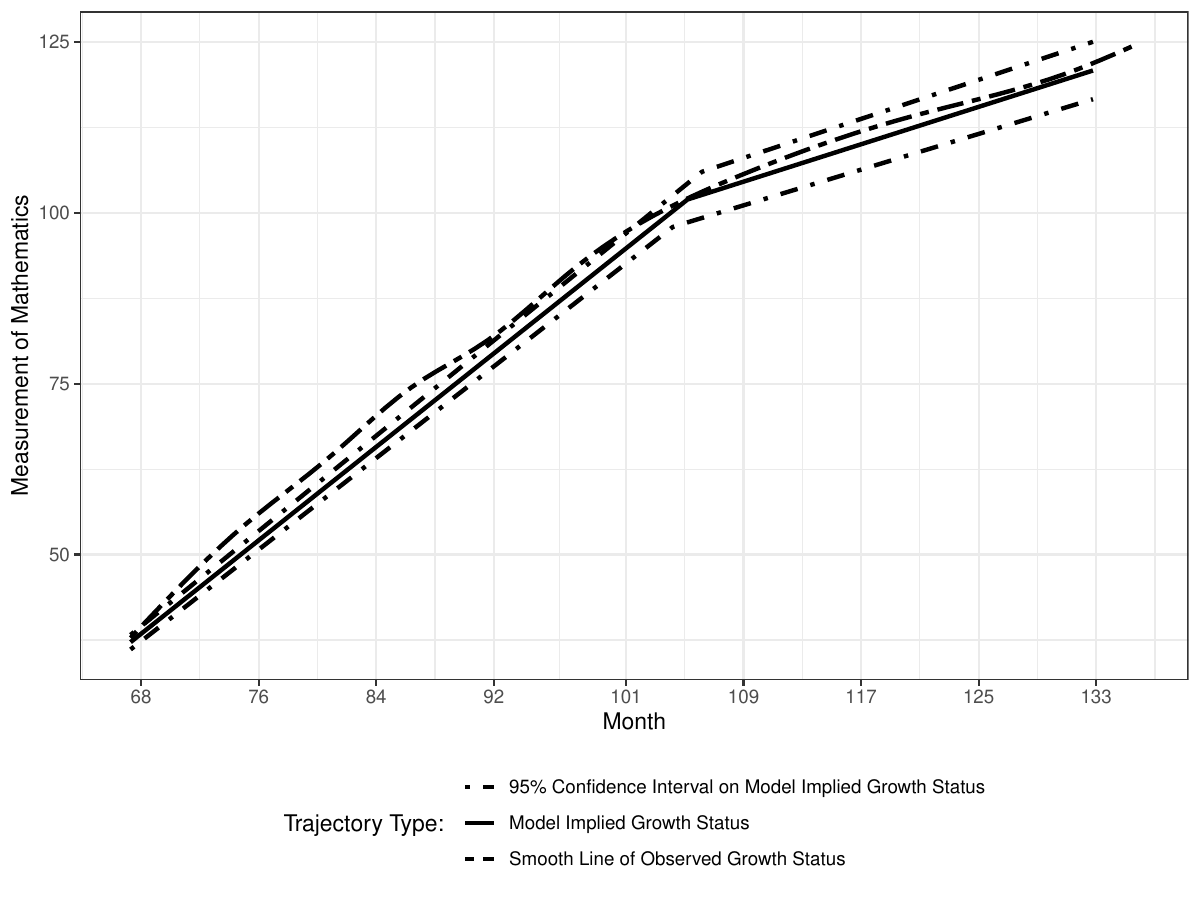}\label{fig:TVC2}}
\caption{Latent Growth Curve Models with Bilinear Spline Function (Random Knots) for Mathematics Ability (TVC: Standardized Reading Ability over Time; Growth TIC: Teacher-reported Approach)}
\label{fig:TVC}
\end{figure}

\begin{figure}[!htbp]
\centering
    \subfloat[Development Status of Reading Ability]{\includegraphics[width=0.5\textwidth]{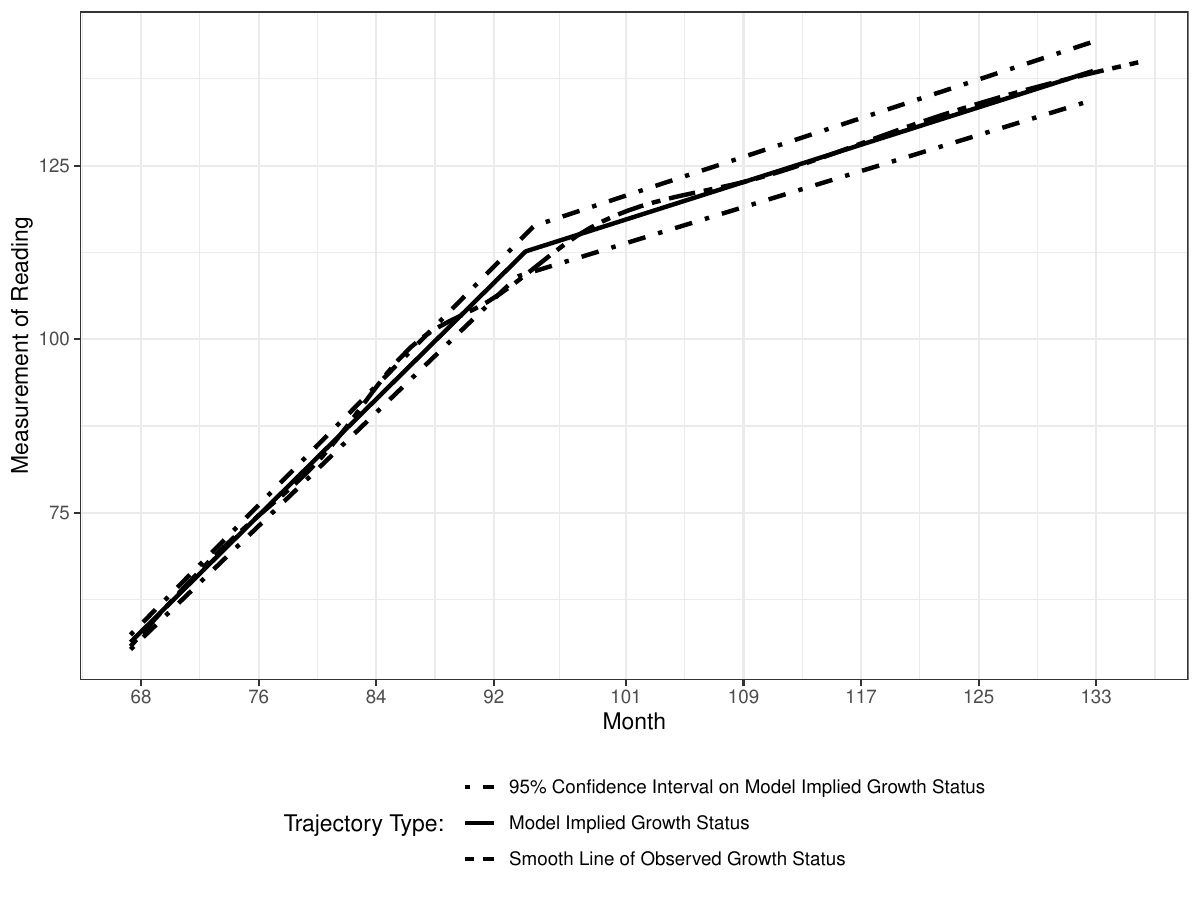}\label{fig:MGM1}}
\hfill
    \subfloat[Development Status of Mathematics Ability]{\includegraphics[width=0.5\textwidth]{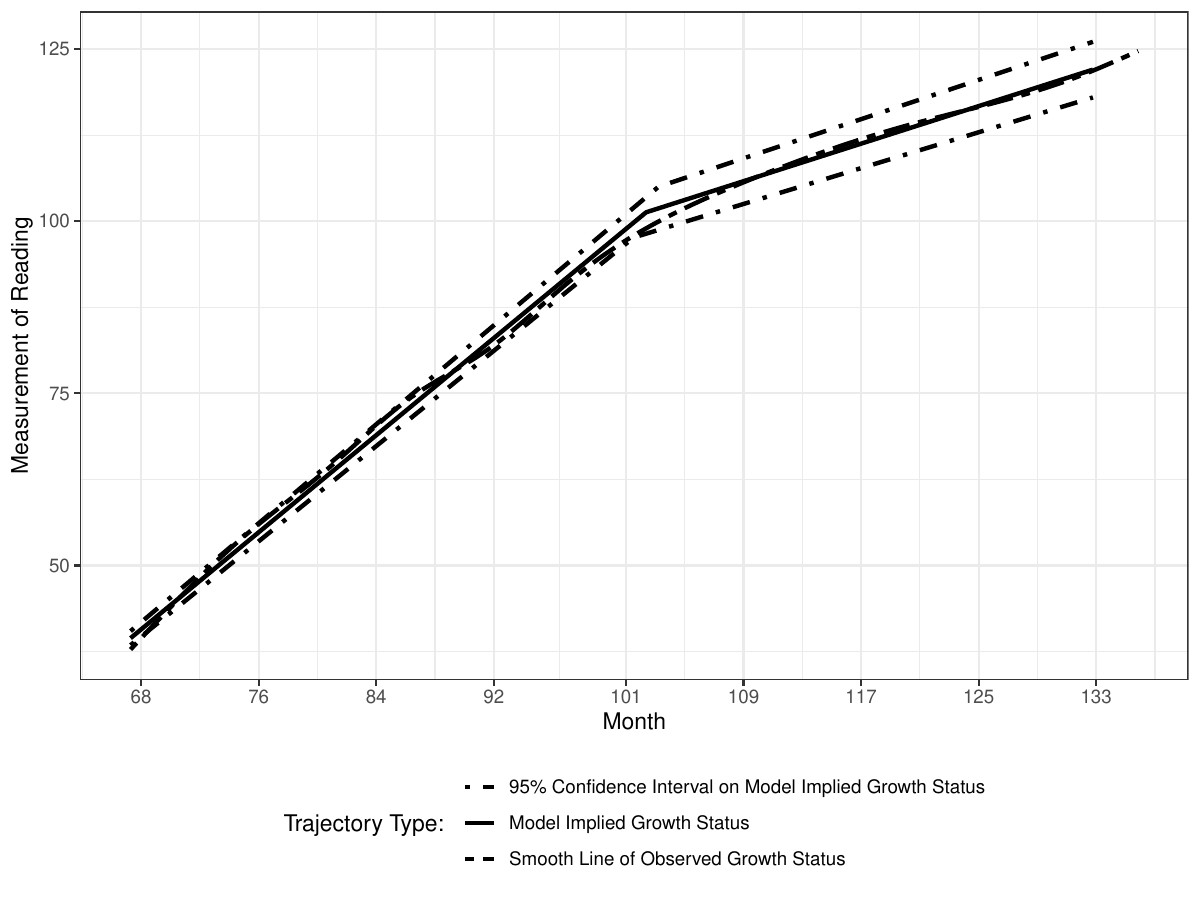}\label{fig:MGM2}}
\caption{Multivariate Latent Growth Curve Models with Bilinear Spline Function (Random Knots) for Reading Ability and Mathematics Ability}
\label{fig:MGM}
\end{figure}

\begin{figure}[!htbp]
\centering
    \subfloat[Mathematics Development by Sex]{\includegraphics[width=0.5\textwidth]{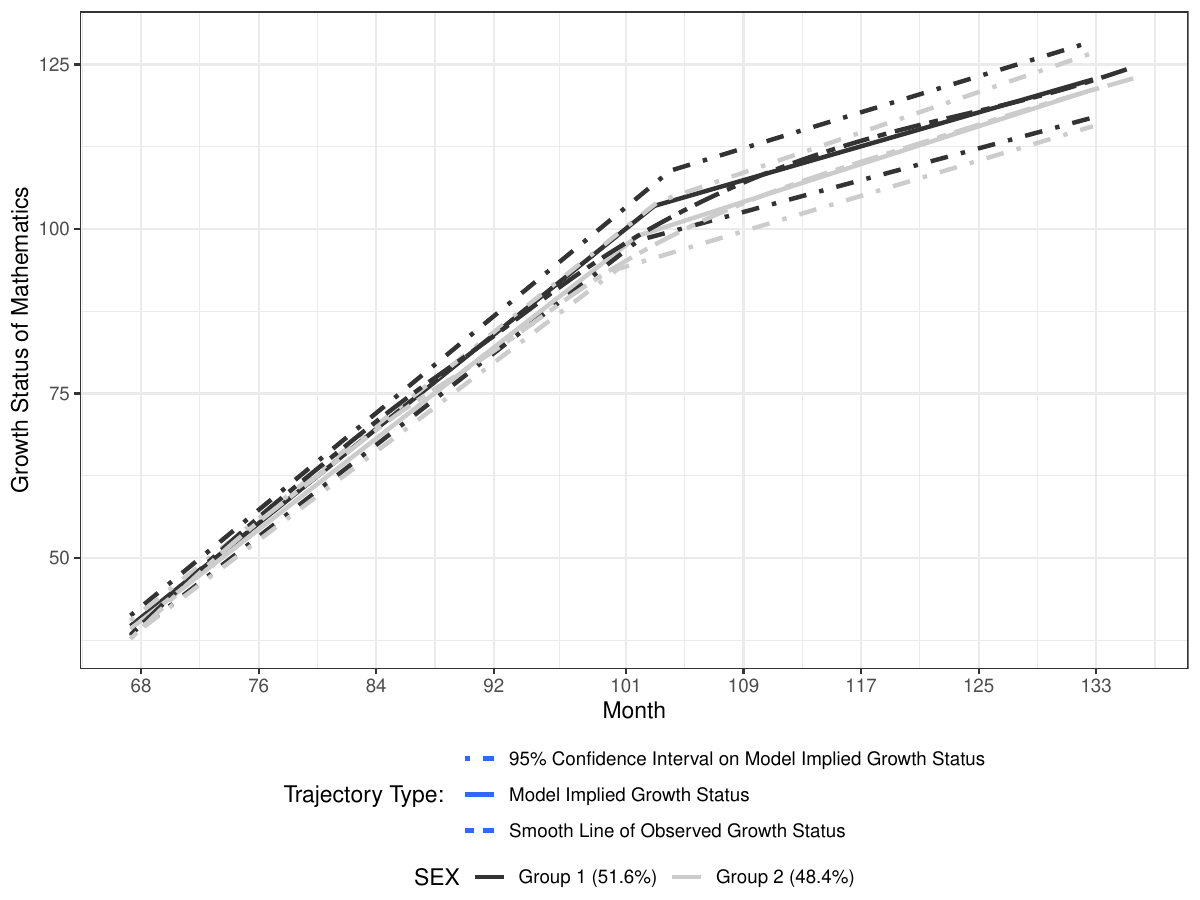}\label{fig:MGroup}}
\hfill
    \subfloat[Mathematics Development by 3 Latent Classes]{\includegraphics[width=0.5\textwidth]{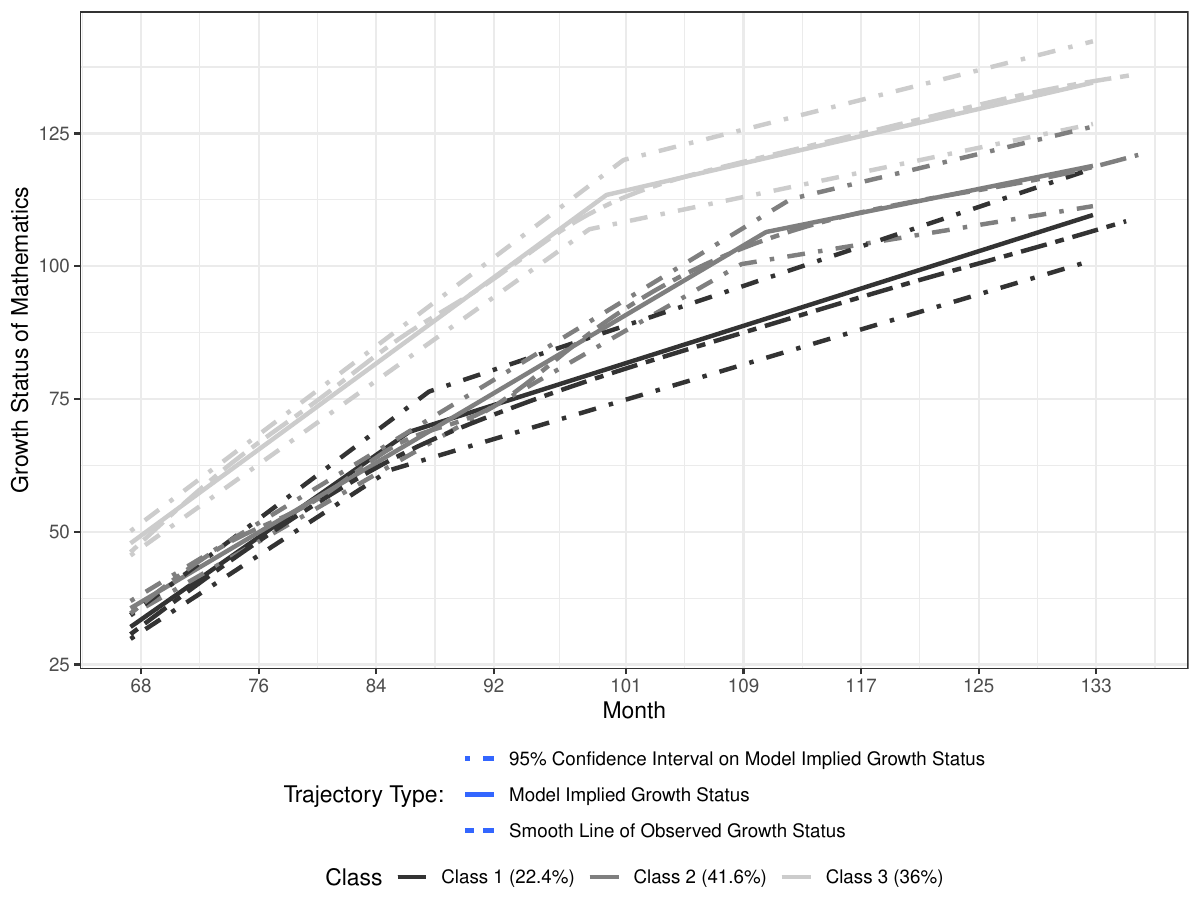}\label{fig:MIX}}
\caption{Multiple Group and Mixture Latent Growth Curve Models with Bilinear Spline Function for Mathematics Ability}
\label{fig:HetGroup}
\end{figure}

\renewcommand\thetable{\arabic{table}}
\setcounter{table}{0}

\begin{table}[!htbp]
\centering
\begin{threeparttable}
\caption{Comparison of Tools Addressing Basic Challenges in Longitudinal Data Analysis}
\begin{tabular}{L{3.8cm}cccccccc}
\hline
\hline
& \pkg{nlme} & \pkg{lme4} & \pkg{lcmm} & \pkg{fitPMM} & \pkg{lavaan} & \pkg{OpenMx} & \pkg{MPlus 8} & \pkg{nlpsem} \\
\hline
SEM Framework & & & & & \checkmark & \checkmark & \checkmark & \checkmark \\
First Type of Nonlinear Models & \checkmark & \checkmark & \checkmark & \checkmark & \checkmark & \checkmark & \checkmark & \checkmark \\
Second Type of Nonlinear Models & \checkmark & \checkmark & \checkmark & \checkmark & & \checkmark & \checkmark & \checkmark \\
Third Type of Nonlinear Models & & & & \checkmark & & \checkmark & \checkmark & \checkmark \\
Allow for Individual-varying Time Points & \checkmark & \checkmark & \checkmark & \checkmark & & \checkmark  & \checkmark & \checkmark \\
User-friendly\tnote{a} & \checkmark & \checkmark & \checkmark & \checkmark & & & & \checkmark \\
\hline
\hline
\end{tabular}
\label{tbl:compare1}
\begin{tablenotes}
\small
\item[a] {User-friendly implies that the tool provides built-in functionalities or templates for common analyses, reducing the need for users to manually specify or script the model. This facilitates a more intuitive and efficient user experience, especially for those who may not be well-versed in the intricacies of model specification.}
\end{tablenotes}
\end{threeparttable}
\end{table}

\begin{table}[!htbp]
\centering
\footnotesize
\resizebox{1.11\columnwidth}{!}{\begin{threeparttable}
\setlength\tabcolsep{2pt}
\renewcommand{\arraystretch}{0.6}
\caption{Model Specification for Commonly Used Latent Growth Curve Models with Individual Measurement Occasions}
\begin{tabular}{p{4.8cm}p{8.7cm}p{6.9cm}}
\hline
\hline
& \multicolumn{2}{c}{\textbf{Linear Function}} \\
\hline
\textbf{Individual Growth Curve}\tnote{a} & $y_{ij}=\eta_{0i}+\eta_{1i}\times{t_{ij}}+\epsilon_{ij}$ & \\
\textbf{Growth Factors}\tnote{b} & $\boldsymbol{\eta}_{i}=\begin{pmatrix}\eta_{0i} & \eta_{1i}\end{pmatrix}$ & \\
\textbf{Factor Loadings}\tnote{b} & $\boldsymbol{\Lambda}_{i}=\begin{pmatrix}1 & t_{ij} \end{pmatrix}$  & \\
\multirow{2}{*}{\textbf{Interpretation of Growth Coef.}} & \multicolumn{2}{l}{$\eta_{0i}$: the individual initial status} \\
& \multicolumn{2}{l}{$\eta_{1i}$: the individual linear component of change} \\
\hline
\hline
& \multicolumn{2}{c}{\textbf{Quadratic Function}} \\
\hline
\textbf{Individual Growth Curve}\tnote{a} & $y_{ij}=\eta_{0i}+\eta_{1i}\times{t_{ij}}+\eta_{2i}\times{t^{2}_{ij}}+\epsilon_{ij}$ & \\
\textbf{Growth Factors}\tnote{b} & $\boldsymbol{\eta}_{i}=\begin{pmatrix}\eta_{0i} & \eta_{1i} & \eta_{2i}\end{pmatrix}$ & \\
\textbf{Factor Loadings}\tnote{b} & $\boldsymbol{\Lambda}_{i}=\begin{pmatrix}1 & t_{ij} & t^{2}_{ij} \end{pmatrix}$  & \\
\multirow{3}{*}{\textbf{Interpretation of Growth Coef.}} & \multicolumn{2}{l}{$\eta_{0i}$: the individual initial status} \\
& \multicolumn{2}{l}{$\eta_{1i}$: the individual linear component of change} \\
& \multicolumn{2}{l}{$\eta_{2i}$: the individual quadratic component of change (i.e., half of the individual growth acceleration)} \\
\hline
\hline
& \multicolumn{2}{c}{\textbf{Negative Exponential Function}} \\
\hline
& \textbf{Intrinsically Nonlinear Model} & 
\textbf{Reduced Non-intrinsically Nonlinear Model} \\
\hline
\textbf{Individual Growth Curve}\tnote{a} & $y_{ij}=\eta_{0i}+\eta_{1i}\times(1-\exp(-b_{i}\times {t_{ij}}))+\epsilon_{ij}$ & $y_{ij}=\eta_{0i}+\eta_{1i}\times(1-\exp(-b\times {t_{ij}}))+\epsilon_{ij}$ \\
\textbf{Growth Factors}\tnote{b} & $\boldsymbol{\eta}_{i}=\begin{pmatrix}\eta_{0i} & \eta_{1i} & b_{i}-\mu_{b} \end{pmatrix}$ & $\boldsymbol{\eta}_{i}=\begin{pmatrix}\eta_{0i} & \eta_{1i} \end{pmatrix}$ \\
\textbf{Factor Loadings}\tnote{b} & $\boldsymbol{\Lambda}_{i}\approx\begin{pmatrix}1 & 1-\exp(-\mu_{b}\times {t_{ij}}) & \mu_{\eta_{1}}\times\exp(-\mu_{b}t_{ij})\times t_{ij} \end{pmatrix}$ & $\boldsymbol{\Lambda}_{i}=\begin{pmatrix}1 & 1-\exp(-b\times {t_{ij}}) \end{pmatrix}$ \\
\multirow{3}{*}{\textbf{Interpretation of Growth Coef.}} & \multicolumn{2}{l}{$\eta_{0i}$: the individual initial status} \\
& \multicolumn{2}{l}{$\eta_{1i}$: the individual change from initial status to asymptotic level (i.e., the individual growth capacity)} \\
& \multicolumn{2}{l}{$b$ ($b_{i}$)\tnote{c}: a growth rate parameter that controls the curvature of the growth trajectory (for individual $i$)} \\
\hline
\hline
& \multicolumn{2}{c}{\textbf{Jenss-Bayley Function}} \\
\hline
& \textbf{Intrinsically Nonlinear Model} & 
\textbf{Reduced Non-intrinsically Nonlinear Model} \\
\hline
\textbf{Individual Growth Curve}\tnote{a} & $y_{ij}=\eta_{0i}+\eta_{1i}\times t_{ij}+\eta_{2i}\times(\exp(c_{i}\times t_{ij})-1)+\epsilon_{ij}$ & $y_{ij}=\eta_{0i}+\eta_{1i}\times t_{ij}+\eta_{2i}\times(\exp(c\times t_{ij})-1)+\epsilon_{ij}$ \\
\textbf{Growth Factors}\tnote{b} & $\boldsymbol{\eta}_{i}=\begin{pmatrix}\eta_{0i} & \eta_{1i} & \eta_{2i} & c_{i}-\mu_{c} \end{pmatrix}$ & $\boldsymbol{\eta}_{i}=\begin{pmatrix}\eta_{0i} & \eta_{1i} & \eta_{2i} \end{pmatrix}$ \\
\textbf{Factor Loadings}\tnote{b} & $\boldsymbol{\Lambda}_{i}\approx\begin{pmatrix}1 & t_{ij} & \exp(\mu_{c}\times {t_{ij}})-1 & \mu_{\eta_{2}}\times\exp(\mu_{c}t_{ij})\times t_{ij} \end{pmatrix}$ & $\boldsymbol{\Lambda}_{i}=\begin{pmatrix}1 & t_{ij} & \exp(c\times {t_{ij}}-1) \end{pmatrix}$ \\
\multirow{4}{*}{\textbf{Interpretation of Growth Coef.}} & \multicolumn{2}{l}{$\eta_{0i}$: the individual initial status} \\
& \multicolumn{2}{l}{$\eta_{1i}$: the individual slope of linear asymptote with the assumption $c_{i}<0$ ($c<0$)\tnote{d}} \\
& \multicolumn{2}{l}{$\eta_{2i}$: the individual change from initial status to the linear asymptote intercept} \\
& \multicolumn{2}{l}{$c$ ($c_{i}$)\tnote{e}: a growth acceleration parameter that controls the rate of change of the growth trajectory's curvature (for individual $i$)} \\
\hline
\hline
& \multicolumn{2}{c}{\textbf{Bilinear Spline Function with an Unknown Knot}} \\
\hline
& \textbf{Intrinsically Nonlinear Model} & 
\textbf{Reduced Non-intrinsically Nonlinear Model} \\
\hline
\textbf{Individual Growth Curve}\tnote{a} & $y_{ij}=\begin{cases}
\eta_{0i}+\eta_{1i}\times t_{ij}+\epsilon_{ij}, &  t_{ij}<\gamma_{i} \\
\eta_{0i}+\eta_{1i}\times \gamma_{i}+\eta_{2i}\times(t_{ij}-\gamma_{i})+\epsilon_{ij}, & t_{ij}\ge\gamma_{i} \\
\end{cases}$ & $y_{ij}=\begin{cases}
\eta_{0i}+\eta_{1i}\times t_{ij}+\epsilon_{ij}, &  t_{ij}<\gamma \\
\eta_{0i}+\eta_{1i}\times \gamma+\eta_{2i}\times(t_{ij}-\gamma)+\epsilon_{ij}, & t_{ij}\ge\gamma \\
\end{cases}$ \\
\textbf{Growth Factors}\tnote{b} & $\boldsymbol{\eta}^{'}_{i}=\begin{pmatrix}\eta_{0i}+\gamma_{i}\eta_{1i} & \frac{\eta_{1i}+\eta_{2i}}{2} & \frac{\eta_{2i}-\eta_{1i}}{2} & \gamma_{i}-\mu_{\gamma}
\end{pmatrix}$ & $\boldsymbol{\eta}^{'}_{i}=\begin{pmatrix}\eta_{0i}+\gamma\eta_{1i} & \frac{\eta_{1i}+\eta_{2i}}{2} & \frac{\eta_{2i}-\eta_{1i}}{2}\end{pmatrix}$ \\
\textbf{Factor Loadings}\tnote{b} & $\boldsymbol{\Lambda}^{'}_{i}\approx\begin{pmatrix}1 & t_{ij}-\mu_{\gamma} & |t_{ij}-\mu_{\gamma}| & -\mu^{'}_{\eta_{2}}-\frac{\mu^{'}_{\eta_{2}}(t_{ij}-\mu_{\gamma})}{|t_{ij}-\mu_{\gamma}|}
\end{pmatrix}$ & $\boldsymbol{\Lambda}^{'}_{i}=\begin{pmatrix}1 & t_{ij}-\gamma & |t_{ij}-\gamma| \end{pmatrix}$ \\
\multirow{4}{*}{\textbf{Interpretation of Growth Coef.}} & \multicolumn{2}{l}{$\eta_{0i}$: the individual initial status} \\
& \multicolumn{2}{l}{$\eta_{1i}$: the individual slope of the first linear piece} \\
& \multicolumn{2}{l}{$\eta_{2i}$: the individual slope of the second linear piece} \\
& \multicolumn{2}{l}{$\gamma$ ($\gamma_{i}$): the (individual) transition time from $1^{st}$ linear piece to $2^{nd}$ linear piece (i.e., knot)} \\
\hline
\hline
\end{tabular}
\label{tbl:LGCM_summary}
\begin{tablenotes}
\small
\item[a] {In the function of the individual growth curve, $y_{ij}$, $t_{ij}$, and $\epsilon_{ij}$ are the observed measurement, recorded time, and residual of the $i^{th}$ individual at the $j^{th}$ time point.} \\
\item[b] {In the vector of growth factors and the corresponding factor loadings, $\mu_{b}$, $\mu_{c}$, and $\mu_{\gamma}$ are the mean values of individual growth rate parameters, of individual growth acceleration parameters, and of individual knots for the negative exponential function, Jenss-Bayley function, and bilinear spline function with an unknown knot, respectively.} \\
\item[c] {There are multiple interpretations for $b$ ($b_{i}$). For example, $\exp(-b_{i}\times(t_{i(j+1)} - t_{ij}))$ represents the ratio of the instantaneous growth rates at time points $t_{i(j+1)}$ and $t_{ij}$. This value reflects how much the growth rate has changed between $t_{ij}$ and $t_{i(j+1)}$, depending on the growth rate parameter $b_{i}$. With the assumption that measurements are taken at equally-spaced waves with scaled intervals, $\exp(b_{i})$ represents the ratio of the instantaneous rates at any adjacent time points.} \\
\item[d] {If $c_{i}>0$ ($c>0$), the Jenss-Bayley function does not have a linear asymptote as the nonlinear component continues to grow with time.} \\
\item[e] {There are multiple interpretations for $c$ ($c_{i}$). For example, $\exp(c_{i}\times(t_{i(j+1)} - t_{ij}))$ represents the ratio of the instantaneous growth accelerations at time points $t_{i(j+1)}$ and $t_{ij}$. This value reflects how much the growth acceleration has changed between $t_{ij}$ and $t_{i(j+1)}$, depending on the growth acceleration parameter $c_{i}$. With the assumption that measurements are taken at equally-spaced waves with scaled intervals, $\exp(c_{i})$ represents the ratio of the instantaneous accelerations at any adjacent time points.}
\end{tablenotes}
\end{threeparttable}}
\end{table}

\begin{table}[!htbp]
\centering
\footnotesize
\resizebox{1.11\columnwidth}{!}{\begin{threeparttable}
\setlength\tabcolsep{2pt}
\renewcommand{\arraystretch}{0.6}
\caption{Model Specification for Commonly Used Latent Change Score Models with Individual Measurement Occasions}
\begin{tabular}{p{5.7cm}p{11.1cm}p{6cm}}
\hline
\hline
& \multicolumn{2}{c}{\textbf{Quadratic Function}} \\
\hline
\textbf{Individual Growth Rate}\tnote{b} & $dy_{ij\_\text{mid}}=\eta_{1i}+2\times\eta_{2i}\times t_{ij\_\text{mid}}$ & \\
\textbf{Growth Factors of Growth Rate}\tnote{c} & $\boldsymbol{\eta}_{di}=\begin{pmatrix}\eta_{1i} & \eta_{2i} \end{pmatrix}$ & \\
\textbf{Factor Loadings of Growth Rate}\tnote{c} & $\boldsymbol{\Lambda}_{di}=\begin{pmatrix} 1 & 2\times t_{ij\_\text{mid}} \end{pmatrix}$ & \\
\textbf{Growth Factors of Growth Status}\tnote{d} & $\boldsymbol{\eta}_{i}=\begin{pmatrix}\eta_{0i} & \eta_{1i} & \eta_{2i}\end{pmatrix}$ & \\
\textbf{Factor Loadings of Growth Status}\tnote{d} & $\boldsymbol{\Lambda}_{i}=\begin{pmatrix} \boldsymbol{1} & \boldsymbol{\Omega}_{i}\times\boldsymbol{\Lambda}_{di}
\end{pmatrix}$  & \\
\hline
\hline
& \multicolumn{2}{c}{\textbf{Negative Exponential Function}} \\
\hline
& \textbf{Intrinsically Nonlinear Model} & 
\textbf{Reduced Non-intrinsically Nonlinear Model} \\
\hline
\textbf{Individual Growth Rate}\tnote{b} & $dy_{ij\_\text{mid}}=b_{i}\times\eta_{1i}\times\exp(-b_{i}\times t_{ij\_\text{mid}})$ & $dy_{ij\_\text{mid}}=b\times\eta_{1i}\times\exp(-b\times t_{ij\_\text{mid}})$ \\
\textbf{Growth Factors of Growth Rate}\tnote{c,e} & $\boldsymbol{\eta}_{di}=\begin{pmatrix}\eta_{1i} & b_{i}-\mu_{b} \end{pmatrix}$ & $\boldsymbol{\eta}_{di}=\eta_{1i}$ \\
\textbf{Factor Loadings of Growth Rate}\tnote{c,e} & $\boldsymbol{\Lambda}_{di}\approx\begin{pmatrix} \mu_{b}\times\exp(-\mu_{b}\times t_{ij\_\text{mid}}) & \mu_{\eta_{1}}\times\exp(-\mu_{b}t_{ij\_\text{mid}})\times(1-\mu_{b}t_{ij\_\text{mid}} \end{pmatrix}$ & $\boldsymbol{\Lambda}_{di}=b\times\exp(-b\times t_{ij\_\text{mid}})$ \\
\textbf{Growth Factors of Growth Status}\tnote{d,e} & $\boldsymbol{\eta}_{i}=\begin{pmatrix}\eta_{0i} & \eta_{1i} & b_{i}-\mu_{b} \end{pmatrix}$ & $\boldsymbol{\eta}_{i}=\begin{pmatrix}\eta_{0i} & \eta_{1i} \end{pmatrix}$ \\
\textbf{Factor Loadings of Growth Status}\tnote{d,e} & $\boldsymbol{\Lambda}_{i}\approx\begin{pmatrix}\boldsymbol{1} & \boldsymbol{\Omega}_{i}\times\boldsymbol{\Lambda}_{di} \end{pmatrix}$ & $\boldsymbol{\Lambda}_{i}=\begin{pmatrix}\boldsymbol{1} & \boldsymbol{\Omega}_{i}\times\boldsymbol{\Lambda}_{di} \end{pmatrix}$ \\
\hline
\hline
& \multicolumn{2}{c}{\textbf{Jenss-Bayley Function}} \\
\hline
& \textbf{Intrinsically Nonlinear Model} & 
\textbf{Reduced Non-intrinsically Nonlinear Model} \\
\hline
\textbf{Individual Growth Rate}\tnote{b} & $dy_{ij\_\text{mid}}=\eta_{1i}+c_{i}\times\eta_{2i}\times\exp(c_{i}\times t_{ij\_\text{mid}})$ & $dy_{ij\_\text{mid}}=\eta_{1i}+c\times\eta_{2i}\times\exp(c\times t_{ij\_\text{mid}})$ \\
\textbf{Growth Factors of Growth Rate}\tnote{c,e} & $\boldsymbol{\eta}_{di}=\begin{pmatrix}\eta_{1i} & \eta_{2i} & c_{i}-\mu_{c} \end{pmatrix}$ & $\boldsymbol{\eta}_{di}=\begin{pmatrix}\eta_{1i} & \eta_{2i} \end{pmatrix}$ \\
\textbf{Factor Loadings of Growth Rate}\tnote{c,e} & $\boldsymbol{\Lambda}_{di}\approx\begin{pmatrix} 1 & \mu_{c}\times\exp(\mu_{c}\times t_{ij\_\text{mid}}) & \mu_{\eta_{2}}\times\exp(\mu_{c}t_{ij\_\text{mid}})\times(1+\mu_{c}t_{ij\_\text{mid}} \end{pmatrix}$ & $\boldsymbol{\Lambda}_{di}\approx\begin{pmatrix} 1 & c\times\exp(c\times t_{ij\_\text{mid}}) \end{pmatrix}$ \\
\textbf{Growth Factors of Growth Status}\tnote{d,e} & $\boldsymbol{\eta}_{i}=\begin{pmatrix}\eta_{0i} & \eta_{1i} & \eta_{2i} & c_{i}-\mu_{c} \end{pmatrix}$ & $\boldsymbol{\eta}_{i}=\begin{pmatrix}\eta_{0i} & \eta_{1i} & \eta_{2i} \end{pmatrix}$ \\
\textbf{Factor Loadings of Growth Status}\tnote{d,e} & $\boldsymbol{\Lambda}_{i}\approx\begin{pmatrix}\boldsymbol{1} & \boldsymbol{\Omega}_{i}\times\boldsymbol{\Lambda}_{di} \end{pmatrix}$ & $\boldsymbol{\Lambda}_{i}=\begin{pmatrix}\boldsymbol{1} & \boldsymbol{\Omega}_{i}\times\boldsymbol{\Lambda}_{di} \end{pmatrix}$ \\
\hline
\hline
& \multicolumn{2}{c}{\textbf{Nonparametric Function}} \\
\hline
\textbf{Individual Growth Rate}\tnote{b} & $dy_{ij}=\eta_{1i}\times\gamma_{j-1}$ & \\
\textbf{Growth Factors of Growth Rate}\tnote{c} & $\boldsymbol{\eta}_{di}=\eta_{1i}$ & \\
\textbf{Factor Loadings of Growth Rate}\tnote{c} & $\boldsymbol{\Lambda}_{di}=\gamma_{j-1}$ & \\
\textbf{Growth Factors of Growth Status}\tnote{d} & $\boldsymbol{\eta}_{i}=\begin{pmatrix}\eta_{0i} & \eta_{1i} \end{pmatrix}$ & \\
\textbf{Factor Loadings of Growth Status}\tnote{d} & $\boldsymbol{\Lambda}_{i}=\begin{pmatrix} \boldsymbol{1} & \boldsymbol{\Omega}_{i}\times\boldsymbol{\Lambda}_{di}
\end{pmatrix}$  & \\
\multirow{3}{*}{\textbf{Interpretation of Growth Coef.}} & \multicolumn{2}{l}{$\eta_{0i}$: the individual initial status} \\
& \multicolumn{2}{l}{$\eta_{1i}$: the individual slope during the first time interval} \\
& \multicolumn{2}{l}{$\gamma_{j}$: the relative growth rate of the $j^{th}$ interval} \\
\hline
\hline
\end{tabular}
\label{tbl:LCSM_summary}
\begin{tablenotes}
\item[a] {This table does not include the specifications for LCSMs with linear and bilinear spline functions, as LGCMs with these two functional forms can estimate interval-specific slopes, eliminating the need for LCSMs to estimate growth rates. Additionally, this table presents the model specifications for the LCSM with a piecewise linear function. Note that the specification of this model serves as the foundation for the models with a decomposed TVC, which is introduced in Subsection \ref{spec:TVC}.} \\
\item[b] {In the individual growth rate function, $dy_{ij\_\text{mid}}$ and $t_{ij\_\text{mid}}$ are the instantaneous slope midway through the $(j-1)^{th}$ time interval and the corresponding time.}\\
\item[c] {The growth factors of the growth rate $\boldsymbol{\eta}_{di}$ consists of those associated with the growth rates, which are present in the respective growth rate function. The corresponding factor loadings are provided in the matrix $\boldsymbol{\Lambda}_{di}$ representing the factor loadings of the growth rate (where $j=2, 3, \dots, J$). The mean vector and variance-covariance matrix of growth factors of growth rate are $\boldsymbol{\mu_{\eta}}_{d}$ and $\boldsymbol{\Psi_{\eta}}_{d}$, respectively. With $\boldsymbol{\eta}_{di}$, $\boldsymbol{\Lambda}_{di}$, $\boldsymbol{\mu_{\eta}}_{d}$, and $\boldsymbol{\Psi_{\eta}}_{d}$, we are able to derive (1) mean and variance of interval-specific slopes: $\boldsymbol{\mu_{dy\_\text{mid}}}=\boldsymbol{\Lambda_{\eta}}_{d}\times\boldsymbol{\mu_{\eta}}_{d}$ and $\boldsymbol{\sigma^{2}_{dy\_\text{mid}}}=\boldsymbol{\Lambda_{\eta}}_{d}\times\boldsymbol{\Psi}_{d}\times\boldsymbol{\Lambda_{\eta}}^{T}_{d}$, and (2) mean and variance of interval-specific changes: $\boldsymbol{\mu_{\delta y_{ij}}}=\boldsymbol{\Lambda_{\eta}}_{d}\times\boldsymbol{\mu_{\eta}}_{d}\times(t_{ij}-t_{i(j-1)})$ and $\boldsymbol{\sigma^{2}_{\delta y_{ij}}}=\boldsymbol{\Lambda_{\eta}}_{d}\times\boldsymbol{\Psi}_{d}\times\boldsymbol{\Lambda_{\eta}}^{T}_{d}\times(t_{ij}-t_{i(j-1)})^2$. In the equations of means and variances of interval-specific slopes and interval-specific changes, $j=2, 3, \dots, J$.}\\
\item[d] {The vector of growth factor of the growth status consists of growth factors associated with both the growth rates and the initial status, which together determine the growth status. The corresponding factor loadings are provided in the matrix representing the factor loadings of the growth status (where $j=1, 2, \dots, J$), in which $\boldsymbol{\Omega}_{i}=\begin{pmatrix}
  0 & 0 & \cdots & \cdots & \cdots & 0 \\
  t_{i2}-t_{i1} & 0 & 0 & \cdots & \cdots & 0 \\
  t_{i2}-t_{i1} & t_{i3}-t_{i2} & 0 & 0 & \cdots & 0 \\
  \cdots & \cdots & \cdots & \cdots & \cdots & \cdots \\
  t_{i2}-t_{i1} & t_{i3}-t_{i2} & t_{i4}-t_{i3} & \cdots & \cdots & t_{ij}-t_{i(j-1)} \\
\end{pmatrix}$ so that $\boldsymbol{\Omega}_{i}\times\boldsymbol{\Lambda}_{di}$ represents the accumulative value since the initial status of the corresponding factor loading of the growth rate. With $\boldsymbol{\Omega}_{i}$, we are able to derive the mean and variance of change from baseline: $\boldsymbol{\mu_{\Delta y_{ij}}}=\boldsymbol{\Omega}_{i}\times\boldsymbol{\Lambda}_{di}\times\boldsymbol{\mu_{\eta}}_{d}$ and $\boldsymbol{\sigma^2_{\Delta y_{ij}}}=\boldsymbol{\Omega}_{i}\times\boldsymbol{\Lambda_{\eta}}_{d}\times\boldsymbol{\Psi}_{d}\times\boldsymbol{\Lambda_{\eta}}^{T}_{d}\times\boldsymbol{\Omega}^{T}_{i}$. In the equations of means and variances of change from baseline, $j=2, 3, \dots, J$.}\\
\item[e] {In the vector of growth factors and the corresponding factor loadings, $\mu_{b}$ and $\mu_{c}$ are the mean values of $b_{i}$ and of $c_{i}$ for the negative exponential function and Jenss-Bayley function, respectively. }
\end{tablenotes}
\end{threeparttable}}
\end{table}

\begin{table}[!htbp]
\centering
\footnotesize
\resizebox{\columnwidth}{!}{\begin{threeparttable}
\setlength\tabcolsep{2pt}
\renewcommand{\arraystretch}{0.6}
\caption{Model Specification for Four Possible Ways of Adding Time-varying Covariate with Individual Measurement Occasions}
\begin{tabular}{p{5cm}p{10cm}}
\hline
\hline
\multicolumn{2}{c}{\textbf{LGCM with a TVC and TICs}} \\
\hline
\multirow{3}{*}{\textbf{Model Specification}} & $\boldsymbol{y}_{i}=\boldsymbol{\Lambda}^{[y]}_{i}\times \boldsymbol{\eta}^{[y]}_{i}+\kappa\times\boldsymbol{x}_{i}+\boldsymbol{\epsilon}^{[y]}_{i}$ \\
& $\boldsymbol{\eta}^{[y]}_{i}=\boldsymbol{\alpha}^{[y]}+\boldsymbol{B}_{\text{TIC}}\times \boldsymbol{X}_{i}+ \boldsymbol{\zeta}^{[y]}_{i}$ \\
\hline
\hline
\multicolumn{2}{c}{\textbf{LGCM with a Decomposed TVC into Baseline and Interval-specific Slopes and TICs}} \\
\hline
\multirow{4}{*}{\textbf{Individual Function of TVC}} & $x_{ij}=x^{\ast}_{ij}+\epsilon^{[x]}_{ij}$ \\
& $x^{\ast}_{ij}=\begin{cases}
\eta^{[x]}_{0i}, & \text{if $j=1$}\\
x^{\ast}_{i(j-1)}+dx_{ij}\times(t_{ij}-t_{i(j-1)}), & \text{if $j=2, \dots, J$}
\end{cases}$ \\
& $dx_{ij}=\eta^{[x]}_{1i}\times\gamma_{j-1}\qquad (j=2, \dots, J)$ \\
\hline
\multirow{6}{*}{\textbf{Model Specification}} & $\begin{pmatrix}\boldsymbol{x}_{i} \\ \boldsymbol{y}_{i}
\end{pmatrix}=\begin{pmatrix}
\boldsymbol{\Lambda}^{[x]}_{i} & \boldsymbol{0} \\
\boldsymbol{0} & \boldsymbol{\Lambda}^{[y]}_{i}
\end{pmatrix}\times\begin{pmatrix} \boldsymbol{\eta}^{[x]}_{i} \\ \boldsymbol{\eta}^{[y]}_{i}
\end{pmatrix}+\kappa\times\begin{pmatrix} \boldsymbol{0} \\ \boldsymbol{dx_{i}}
\end{pmatrix}+\begin{pmatrix}
\boldsymbol{\epsilon}^{[x]}_{i} \\ \boldsymbol{\epsilon}^{[y]}_{i}
\end{pmatrix}$ \\
& $\boldsymbol{x}_{i}=\boldsymbol{\Lambda}^{[x]}_{i}\times\boldsymbol{\eta}^{[x]}_{i}+\boldsymbol{\epsilon}^{[x]}_{i}$ \\
& $\boldsymbol{\eta}^{[y]}_{i}=\boldsymbol{\alpha}^{[y]}+\begin{pmatrix}\boldsymbol{B}_{\text{TIC}} & \boldsymbol{\beta}_{\text{TVC}}\end{pmatrix}\times\begin{pmatrix}\boldsymbol{X}_{i} \\ \eta^{[x]}_{0i}\end{pmatrix} +\boldsymbol{\zeta}^{[y]}_{i}$ \\
& $\boldsymbol{dx_{i}}=\begin{pmatrix}0 & dx_{i2} & dx_{i3} & \dots & dx_{iJ}\end{pmatrix}^{T}$ \\
\hline
\hline
\multicolumn{2}{c}{\textbf{LGCM with a Decomposed TVC into Baseline and Interval-specific Changes and TICs}} \\
\hline
\multirow{5}{*}{\textbf{Individual Function of TVC}} & $x_{ij}=x^{\ast}_{ij}+\epsilon^{[x]}_{ij}$ \\
& $x^{\ast}_{ij}=\begin{cases}
\eta^{[x]}_{0i}, & \text{if $j=1$}\\
x^{\ast}_{i(j-1)}+\delta x_{ij}, & \text{if $j=2, \dots, J$}
\end{cases}$ \\
& $\delta x_{ij}=dx_{ij}\times(t_{ij}-t_{i(j-1)})\qquad (j=2, \dots, J)$ \\
& $dx_{ij}=\eta^{[x]}_{1i}\times\gamma_{j-1}\qquad (j=2, \dots, J)$ \\
\hline
\multirow{6}{*}{\textbf{Model Specification}} & $\begin{pmatrix}\boldsymbol{x}_{i} \\ \boldsymbol{y}_{i}
\end{pmatrix}=\begin{pmatrix}
\boldsymbol{\Lambda}^{[x]}_{i} & \boldsymbol{0} \\
\boldsymbol{0} & \boldsymbol{\Lambda}^{[y]}_{i}
\end{pmatrix}\times\begin{pmatrix} \boldsymbol{\eta}^{[x]}_{i} \\ \boldsymbol{\eta}^{[y]}_{i}
\end{pmatrix}+\kappa\times\begin{pmatrix} \boldsymbol{0} \\ \boldsymbol{\delta x_{i}}
\end{pmatrix}+\begin{pmatrix}
\boldsymbol{\epsilon}^{[x]}_{i} \\ \boldsymbol{\epsilon}^{[y]}_{i}
\end{pmatrix}$ \\
& $\boldsymbol{x}_{i}=\boldsymbol{\Lambda}^{[x]}_{i}\times\boldsymbol{\eta}^{[x]}_{i}+\boldsymbol{\epsilon}^{[x]}_{i}$ \\
& $\boldsymbol{\eta}^{[y]}_{i}=\boldsymbol{\alpha}^{[y]}+\begin{pmatrix}\boldsymbol{B}_{\text{TIC}} & \boldsymbol{\beta}_{\text{TVC}}\end{pmatrix}\times\begin{pmatrix}\boldsymbol{X}_{i} \\ \eta^{[x]}_{0i}\end{pmatrix} +\boldsymbol{\zeta}^{[y]}_{i}$ \\
& $\boldsymbol{\delta x_{i}}=\begin{pmatrix}0 & \delta x_{i2} & \delta x_{i3} & \dots & \delta x_{iJ}\end{pmatrix}^{T}$ \\
\hline
\hline
\multicolumn{2}{c}{\textbf{LGCM with a Decomposed TVC into Baseline and Change-from-baseline and TICs}} \\
\hline
\multirow{5}{*}{\textbf{Individual Function of TVC}} & $x_{ij}=x^{\ast}_{ij}+\epsilon^{[x]}_{ij}$ \\
& $x^{\ast}_{ij}=\begin{cases}
\eta^{[x]}_{0i}, & \text{if $j=1$}\\
\eta^{[x]}_{0i}+\Delta x_{ij}, & \text{if $j=2, \dots, J$}
\end{cases}$ \\
& $\Delta x_{ij}=\Delta x_{i(j-1)}+dx_{ij}\times(t_{ij}-t_{i(j-1)})$ \\
& $dx_{ij}=\eta^{[x]}_{1i}\times\gamma_{j-1}\qquad (j=2, \dots, J)$ \\
\hline
\multirow{6}{*}{\textbf{Model Specification}} & $\begin{pmatrix}\boldsymbol{x}_{i} \\ \boldsymbol{y}_{i}
\end{pmatrix}=\begin{pmatrix}
\boldsymbol{\Lambda}^{[x]}_{i} & \boldsymbol{0} \\
\boldsymbol{0} & \boldsymbol{\Lambda}^{[y]}_{i}
\end{pmatrix}\times\begin{pmatrix} \boldsymbol{\eta}^{[x]}_{i} \\ \boldsymbol{\eta}^{[y]}_{i}
\end{pmatrix}+\kappa\times\begin{pmatrix} \boldsymbol{0} \\ \boldsymbol{\Delta x_{i}}
\end{pmatrix}+\begin{pmatrix}
\boldsymbol{\epsilon}^{[x]}_{i} \\ \boldsymbol{\epsilon}^{[y]}_{i}
\end{pmatrix}$ \\
& $\boldsymbol{x}_{i}=\boldsymbol{\Lambda}^{[x]}_{i}\times\boldsymbol{\eta}^{[x]}_{i}+\boldsymbol{\epsilon}^{[x]}_{i}$ \\
& $\boldsymbol{\eta}^{[y]}_{i}=\boldsymbol{\alpha}^{[y]}+\begin{pmatrix}\boldsymbol{B}_{\text{TIC}} & \boldsymbol{\beta}_{\text{TVC}}\end{pmatrix}\times\begin{pmatrix}\boldsymbol{X}_{i} \\ \eta^{[x]}_{0i}\end{pmatrix} +\boldsymbol{\zeta}^{[y]}_{i}$ \\
& $\boldsymbol{\Delta x_{i}}=\begin{pmatrix}0 & \Delta x_{i2} & \Delta x_{i3} & \dots & \Delta x_{iJ}\end{pmatrix}^{T}$ \\
\hline
\hline
\end{tabular}
\label{tbl:TVC_summary}
\end{threeparttable}}
\end{table}

\begin{table}[!htbp]
\centering
\footnotesize
\resizebox{0.95\columnwidth}{!}{\begin{threeparttable}
\setlength\tabcolsep{2pt}
\renewcommand{\arraystretch}{0.6}
\caption{Model Specification for Longitudinal Mediation Models with Individual Measurement Occasions}
\begin{tabular}{p{3cm}p{4cm}p{12.6cm}}
\hline
\hline
\multicolumn{3}{c}{\textbf{Baseline Covariate, Longitudinal Mediator, and Longitudinal Outcome}} \\
\hline
\multirow{10}{*}{\textbf{Linear Function}} & \multirow{6}{*}{\textbf{Model Specification}} & $\begin{pmatrix}
\boldsymbol{m}_{i} \\ \boldsymbol{y}_{i}
\end{pmatrix}=
\begin{pmatrix}
\boldsymbol{\Lambda}_{i}^{[m]} & \boldsymbol{0} \\ \boldsymbol{0} & \boldsymbol{\Lambda}_{i}^{[y]}
\end{pmatrix}\times
\begin{pmatrix}
\boldsymbol{\eta}^{[m]}_{i} \\ \boldsymbol{\eta}^{[y]}_{i}
\end{pmatrix}+
\begin{pmatrix}
\boldsymbol{\epsilon}^{[m]}_{i} \\ \boldsymbol{\epsilon}^{[y]}_{i}
\end{pmatrix}$ \\
& & $\boldsymbol{\eta}^{[u]}_{i} = \begin{pmatrix}
\eta^{[u]}_{0i} & \eta^{[u]}_{1i} \end{pmatrix}^{T}$ $(u=m,y)$ \\
& & $\boldsymbol{\Lambda}^{[u]}_{i} = \begin{pmatrix}
0 & t_{ij} \end{pmatrix}$ $(u=m,y; j=1,\cdots, J)$ \\
& & $\boldsymbol{\eta}^{[m]}_{i}=\boldsymbol{\alpha}^{[m]}+\boldsymbol{B}^{[x\rightarrow{m}]}\times x_{i}+\boldsymbol{\zeta}^{[m]}_{i}$ \\
& & $\boldsymbol{\eta}^{[y]}_{i}=\boldsymbol{\alpha}^{[y]}+\boldsymbol{B}^{[x\rightarrow{y}]}\times x_{i}+\boldsymbol{B}^{[m\rightarrow{y}]}\times\boldsymbol{\eta}^{[m]}_{i}+\boldsymbol{\zeta}^{[y]}_{i}$ \\
\cline{2-3} 
& \textbf{Growth Factor Intercept} & $\boldsymbol{\alpha}^{[u]}_{i} = \begin{pmatrix}
\alpha^{[u]}_{0i} & \alpha^{[u]}_{1i} \end{pmatrix}^{T}$ $(u=m,y)$ \\
\cline{2-3}
& \multirow{4}{*}{\textbf{Path Coef.}} & $\boldsymbol{B}^{[x\rightarrow{m}]}=\begin{pmatrix}
\beta^{[x\rightarrow{m}]}_{0} & \beta^{[x\rightarrow{m}]}_{1} \end{pmatrix}^{T}$; $\boldsymbol{B}^{[x\rightarrow{y}]}=\begin{pmatrix}
\beta^{[x\rightarrow{y}]}_{0} & \beta^{[x\rightarrow{y}]}_{1} \end{pmatrix}^{T}$ \\
& & $\boldsymbol{B}^{[m\rightarrow{y}]}=\begin{pmatrix}
\beta^{[m\rightarrow{y}]}_{00} & 0 \\
\beta^{[m\rightarrow{y}]}_{01} & \beta^{[m\rightarrow{y}]}_{11} \\
\end{pmatrix}$ \\
\hline
\hline
\multirow{11}{*}{\textbf{Bilinear Function}} & \multirow{6}{*}{\textbf{Model Specification}} & $\begin{pmatrix}
\boldsymbol{m}_{i} \\ \boldsymbol{y}_{i}
\end{pmatrix}=
\begin{pmatrix}
\boldsymbol{\Lambda}_{i}^{[m]} & \boldsymbol{0} \\ \boldsymbol{0} & \boldsymbol{\Lambda}_{i}^{[y]}
\end{pmatrix}\times
\begin{pmatrix}
\boldsymbol{\eta}^{[m]}_{i} \\ \boldsymbol{\eta}^{[y]}_{i}
\end{pmatrix}+
\begin{pmatrix}
\boldsymbol{\epsilon}^{[m]}_{i} \\ \boldsymbol{\epsilon}^{[y]}_{i}
\end{pmatrix}$ \\
& & $\boldsymbol{\eta}^{[u]}_{i} = \begin{pmatrix}
\eta^{[u]}_{1i} & \eta^{[u]}_{\gamma_{i}} & \eta^{[u]}_{2i} 
\end{pmatrix}^{T}$ $(u=m,y)$ \\
& & $\boldsymbol{\Lambda}^{[u]}_{i} = \begin{pmatrix}
\min(0,t_{ij}-\gamma^{[u]}) & 1 & \max(0,t_{ij}-\gamma^{[u]})
\end{pmatrix}$ $(u=m,y; j=1,\cdots, J)$ \\
& & $\boldsymbol{\eta}^{[m]}_{i}=\boldsymbol{\alpha}^{[m]}+\boldsymbol{B}^{[x\rightarrow{m}]}\times x_{i}+\boldsymbol{\zeta}^{[m]}_{i}$ \\
& & $\boldsymbol{\eta}^{[y]}_{i}=\boldsymbol{\alpha}^{[y]}+\boldsymbol{B}^{[x\rightarrow{y}]}\times x_{i}+\boldsymbol{B}^{[m\rightarrow{y}]}\times\boldsymbol{\eta}^{[m]}_{i}+\boldsymbol{\zeta}^{[y]}_{i}$ \\
\cline{2-3}
& \textbf{Growth Factor Intercept} & $\boldsymbol{\alpha}^{[u]}_{i} = \begin{pmatrix}
\alpha^{[u]}_{1i} & \alpha^{[u]}_{\gamma_{i}} & \alpha^{[u]}_{2i} 
\end{pmatrix}^{T}$ $(u=m,y)$ \\
\cline{2-3}
& \multirow{4}{*}{\textbf{Path Coef.}} & $\boldsymbol{B}^{[x\rightarrow{m}]}=\begin{pmatrix}
\beta^{[x\rightarrow{m}]}_{1} & \beta^{[x\rightarrow{m}]}_{\gamma} & \beta^{[x\rightarrow{m}]}_{2}
\end{pmatrix}^{T}$; $\boldsymbol{B}^{[x\rightarrow{y}]}=\begin{pmatrix}
\beta^{[x\rightarrow{y}]}_{1} & \beta^{[x\rightarrow{y}]}_{\gamma} & \beta^{[x\rightarrow{y}]}_{2}
\end{pmatrix}^{T}$ \\
& & $\boldsymbol{B}^{[m\rightarrow{y}]}=\begin{pmatrix}
\beta^{[m\rightarrow{y}]}_{11} & 0 & 0 \\
\beta^{[m\rightarrow{y}]}_{1\gamma} & \beta^{[m\rightarrow{y}]}_{\gamma\gamma} & 0 \\
\beta^{[m\rightarrow{y}]}_{12} & \beta^{[m\rightarrow{y}]}_{\gamma2} & \beta^{[m\rightarrow{y}]}_{22} \\
\end{pmatrix}$ \\
\hline
\hline
\multicolumn{3}{c}{\textbf{Longitudinal Covariate, Longitudinal Mediator, and Longitudinal Outcome}} \\
\hline
\multirow{17}{*}{\textbf{Linear Function}} & \multirow{8}{*}{\textbf{Model Specification}} & $\begin{pmatrix}
\boldsymbol{x}_{i} \\ \boldsymbol{m}_{i} \\ \boldsymbol{y}_{i}
\end{pmatrix}=
\begin{pmatrix}
\boldsymbol{\Lambda}_{i}^{[x]} & \boldsymbol{0} & \boldsymbol{0} \\
\boldsymbol{0} & \boldsymbol{\Lambda}_{i}^{[m]} & \boldsymbol{0} \\ 
\boldsymbol{0} & \boldsymbol{0} & \boldsymbol{\Lambda}_{i}^{[y]}
\end{pmatrix}\times
\begin{pmatrix}
\boldsymbol{\eta}^{[x]}_{i} \\\boldsymbol{\eta}^{[m]}_{i} \\ \boldsymbol{\eta}^{[y]}_{i}
\end{pmatrix}+
\begin{pmatrix}
\boldsymbol{\epsilon}^{[x]}_{i} \\\boldsymbol{\epsilon}^{[m]}_{i} \\ \boldsymbol{\epsilon}^{[y]}_{i}
\end{pmatrix}$ \\
& & $\boldsymbol{\eta}^{[u]}_{i} = \begin{pmatrix}
\eta^{[u]}_{0i} & \eta^{[u]}_{1i} \end{pmatrix}^{T}$ $(u=x,m,y)$ \\
& & $\boldsymbol{\Lambda}^{[u]}_{i} = \begin{pmatrix}
1 & t_{ij} \end{pmatrix}$ $(u=x,m,y; j=1,\cdots, J)$ \\
& & $\boldsymbol{\eta}^{[x]}_{i}=\boldsymbol{\mu}^{[x]}_{\boldsymbol{\eta}}+\boldsymbol{\zeta}^{[x]}_{i}$ \\
& & $\boldsymbol{\eta}^{[m]}_{i}=\boldsymbol{\alpha}^{[m]}+\boldsymbol{B}^{[x\rightarrow{m}]}\times\boldsymbol{\eta}^{[x]}_{i}+\boldsymbol{\zeta}^{[m]}_{i}$ \\
& & $\boldsymbol{\eta}^{[y]}_{i}=\boldsymbol{\alpha}^{[y]}+\boldsymbol{B}^{[x\rightarrow{y}]}\times\boldsymbol{\eta}^{[x]}_{i}+\boldsymbol{B}^{[m\rightarrow{y}]}\times\boldsymbol{\eta}^{[m]}_{i}+\boldsymbol{\zeta}^{[y]}_{i}$ \\
\cline{2-3}
& \textbf{Growth Factor Intercept} & $\boldsymbol{\alpha}^{[u]}_{i} = \begin{pmatrix}
\alpha^{[u]}_{0i} & \alpha^{[u]}_{1i} \end{pmatrix}^{T}$ $(u=m,y)$ \\
\cline{2-3}
& \multirow{4}{*}{\textbf{Path Coef.}} & $\boldsymbol{B}^{[x\rightarrow{m}]}=\begin{pmatrix}
\beta^{[x\rightarrow{m}]}_{00} & 0 \\
\beta^{[x\rightarrow{m}]}_{01} & \beta^{[x\rightarrow{m}]}_{11} \end{pmatrix}$; $\boldsymbol{B}^{[x\rightarrow{y}]}=\begin{pmatrix}
\beta^{[x\rightarrow{y}]}_{00} & 0 \\
\beta^{[x\rightarrow{y}]}_{01} & \beta^{[x\rightarrow{y}]}_{11} \\
\end{pmatrix}$ \\
& & $\boldsymbol{B}^{[m\rightarrow{y}]}=\begin{pmatrix}
\beta^{[m\rightarrow{y}]}_{00} & 0 \\
\beta^{[m\rightarrow{y}]}_{01} & \beta^{[m\rightarrow{y}]}_{11} \end{pmatrix}$ \\
\hline
\hline
\multicolumn{3}{c}{\textbf{Longitudinal Covariate, Longitudinal Mediator, and Longitudinal Outcome}} \\
\hline
\multirow{17}{*}{\textbf{Bilinear Function}} & \multirow{8}{*}{\textbf{Model Specification}} & $\begin{pmatrix}
\boldsymbol{x}_{i} \\ \boldsymbol{m}_{i} \\ \boldsymbol{y}_{i}
\end{pmatrix}=
\begin{pmatrix}
\boldsymbol{\Lambda}_{i}^{[x]} & \boldsymbol{0} & \boldsymbol{0} \\
\boldsymbol{0} & \boldsymbol{\Lambda}_{i}^{[m]} & \boldsymbol{0} \\ 
\boldsymbol{0} & \boldsymbol{0} & \boldsymbol{\Lambda}_{i}^{[y]}
\end{pmatrix}\times
\begin{pmatrix}
\boldsymbol{\eta}^{[x]}_{i} \\\boldsymbol{\eta}^{[m]}_{i} \\ \boldsymbol{\eta}^{[y]}_{i}
\end{pmatrix}+
\begin{pmatrix}
\boldsymbol{\epsilon}^{[x]}_{i} \\\boldsymbol{\epsilon}^{[m]}_{i} \\ \boldsymbol{\epsilon}^{[y]}_{i}
\end{pmatrix}$ \\
& & $\boldsymbol{\eta}^{[u]}_{i} = \begin{pmatrix}
\eta^{[u]}_{1i} & \eta^{[u]}_{\gamma_{i}} & \eta^{[u]}_{2i} 
\end{pmatrix}^{T}$ $(u=x,m,y)$ \\
& & $\boldsymbol{\Lambda}^{[u]}_{i} = \begin{pmatrix}
\min(0,t_{ij}-\gamma^{[u]}) & 1 & \max(0,t_{ij}-\gamma^{[u]})
\end{pmatrix}$ $(u=x,m,y; j=1,\cdots, J)$ \\
& & $\boldsymbol{\eta}^{[x]}_{i}=\boldsymbol{\mu}^{[x]}_{\boldsymbol{\eta}}+\boldsymbol{\zeta}^{[x]}_{i}$ \\
& & $\boldsymbol{\eta}^{[m]}_{i}=\boldsymbol{\alpha}^{[m]}+\boldsymbol{B}^{[x\rightarrow{m}]}\times\boldsymbol{\eta}^{[x]}_{i}+\boldsymbol{\zeta}^{[m]}_{i}$ \\
& & $\boldsymbol{\eta}^{[y]}_{i}=\boldsymbol{\alpha}^{[y]}+\boldsymbol{B}^{[x\rightarrow{y}]}\times\boldsymbol{\eta}^{[x]}_{i}+\boldsymbol{B}^{[m\rightarrow{y}]}\times\boldsymbol{\eta}^{[m]}_{i}+\boldsymbol{\zeta}^{[y]}_{i}$ \\
\cline{2-3}
& \textbf{Growth Factor Intercept} & $\boldsymbol{\alpha}^{[u]}_{i} = \begin{pmatrix}
\alpha^{[u]}_{1i} & \alpha^{[u]}_{\gamma_{i}} & \alpha^{[u]}_{2i} 
\end{pmatrix}^{T}$ $(u=m,y)$ \\
\cline{2-3}
& \multirow{6}{*}{\textbf{Path Coef.}} & $\boldsymbol{B}^{[x\rightarrow{m}]}=\begin{pmatrix}
\beta^{[x\rightarrow{m}]}_{11} & 0 & 0 \\
\beta^{[x\rightarrow{m}]}_{1\gamma} & \beta^{[x\rightarrow{m}]}_{\gamma\gamma} & 0 \\
\beta^{[x\rightarrow{m}]}_{12} & \beta^{[x\rightarrow{m}]}_{\gamma2} & \beta^{[x\rightarrow{m}]}_{22} \\
\end{pmatrix}$; $\boldsymbol{B}^{[x\rightarrow{y}]}=\begin{pmatrix}
\beta^{[x\rightarrow{y}]}_{11} & 0 & 0 \\
\beta^{[x\rightarrow{y}]}_{1\gamma} & \beta^{[x\rightarrow{y}]}_{\gamma\gamma} & 0 \\
\beta^{[x\rightarrow{y}]}_{12} & \beta^{[x\rightarrow{y}]}_{\gamma2} & \beta^{[x\rightarrow{y}]}_{22} \\
\end{pmatrix}$ \\
& & $\boldsymbol{B}^{[m\rightarrow{y}]}=\begin{pmatrix}
\beta^{[m\rightarrow{y}]}_{11} & 0 & 0 \\
\beta^{[m\rightarrow{y}]}_{1\gamma} & \beta^{[m\rightarrow{y}]}_{\gamma\gamma} & 0 \\
\beta^{[m\rightarrow{y}]}_{12} & \beta^{[m\rightarrow{y}]}_{\gamma2} & \beta^{[m\rightarrow{y}]}_{22} 
\end{pmatrix}$ \\
\hline
\hline
\end{tabular}
\label{tbl:Med_summary}
\end{threeparttable}}
\end{table}

\end{document}